\documentclass[10pt,twocolumn,prb,aps,amsmath,amssymb,superscriptaddress]{revtex4-2}
\usepackage{graphicx}% Include figure files
\usepackage{dcolumn}% Align table columns on decimal point
\usepackage{bm}% bold math
\usepackage{float}
\usepackage[colorlinks=true,linkcolor=blue,citecolor=blue,urlcolor=blue]{hyperref}
\usepackage{natbib}
\usepackage{color}

\begin{document}

%\preprint{APS/123-QED}

% \title{Extension of the Stoner–Wohlfarth model to MNPs with variable shape and elongation}
\title{Including nanoparticle shape into macrospin models}
%\title{Extending the Stoner–Wohlfarth model to non-spherical and elongated nanoparticles}
%\title{Extending the Stoner--Wohlfarth model to particles with different shapes and elongations}
%\title{Extending the Stoner--Wohlfarth model to particle shapes and elongations}

\author{Iago López-Vázquez}
\email{iago.lopez@usc.es}
\affiliation{Dpt. de Física Aplicada, Universidade de Santiago de Compostela, 15782 Santiago de Compostela, Galicia, Spain}
\affiliation{Instituto de Materiais (iMATUS), Universidade de Santiago de Compostela, 15782 Santiago de Compostela, Galicia, Spain}

\author{Òscar Iglesias}
\email{oscariglesias@ub.edu}
\affiliation{Dpt. de Física de la Matèria Condensada, Universitat de Barcelona and IN2UB, c/ Mart\'{\i} i Franquès 1, 08028 Barcelona, Catalunya, Spain}

\author{David Serantes}
\email{david.serantes@usc.gal}
\affiliation{Dpt. de Física Aplicada, Universidade de Santiago de Compostela, 15782 Santiago de Compostela, Galicia, Spain}
\affiliation{Instituto de Materiais (iMATUS), Universidade de Santiago de Compostela, 15782 Santiago de Compostela, Galicia, Spain}

\date{\today}

\begin{abstract}
We investigate the feasibility of the macrospin approximation to account for the actual shape of soft magnetic nanoparticles (MNPs) with realistic geometries. Specifically focusing on magnetite, we use the superellipsoidal parametrisation to account for a variety of shapes, with a continuous interpolation from spherical to cubic morphologies, as well as different elongations. Our procedure consists of the direct comparison between angular-dependent hysteresis loops obtained by full micromagnetic simulations, with those produced by an extended Stoner–Wohlfarth (SW) model that incorporates both the intrinsic cubic magnetocrystalline anisotropy, and an effective uniaxial contribution arising from the particle elongation. Our results show that the extended SW framework provides quantitative agreement with micromagnetics over a broad range of particle volumes and aspect ratios, demonstrating that the effective uniaxial term captures the dominant shape‑induced contributions. The limits of validity of the macrospin description are approximately 10–60 nm for axial ratios $r>1.5$, and 20–60 nm for $1.0 < r<1.5$. These results establish a direct connection between nanoparticle morphology and effective macrospin parameters, demonstrating the suitability of the generalized SW model for describing the magnetic response of realistically shaped MNPs.

\end{abstract}

\maketitle

%\tableofcontents

\section{Introduction}

%Magnetic nanoparticles (MNPs) have received huge research attention for biomedical applications, mainly based on their good biocompatibility and the possibility to trigger their response remotely by harmless magnetic fields. Promising applications would range from hyperthermia cancer treatment to enhanced imaging or localised drug release. 

%However, in spite of the efforts made, the success to date has been very scarce, with several factors hindering its widespread use that range from delivery challenges [REF] to difficulties in achieving precise remote manipulation [REF]. From a fundamental physics standpoint, we identify an additional issue: nanoparticles models are rather limited to describe the real particle systems.

% \CommDavid{\textcolor{ref}{ÒSCAR, cómo se relaciona el artículo este de Garanin con lo nuestro...? https://arxiv.org/abs/1803.10406 Digo en relación a lo que dice de límites de additive macrospin models... (pregunta también para ti, Iago!)}}

The theory developed by E. C. Stoner and E. P. Wohlfarth \cite{Stoner1948} remains a cornerstone for understanding the behavior of magnetic nanoparticle (MNP) systems \cite{tannous2008stoner-STATIC,tannous2008stoner-DYNAMIC,doyle2009stoner}. Their key aspect was to consider the magnetization of the particle to remain essentially uniform under applied and/or thermal fields. This implies that the particle size and shape lie within the single-domain range, so that the magnetization rotates coherently. In this regime, the particle’s magnetic moment can be regarded as a large macrospin. This formalism has enabled the development of advanced theoretical approaches to describe the magnetization dynamics of nanoparticle ensembles \cite{pfeiffer1990determination,Chantrell2000,ruta2015unified}.

A major advantage of the SW model for interpreting experimental data is the direct correlation between the intrinsic particle parameters—saturation magnetization ($M_S$) and anisotropy constant ($K$)—and measurable macroscopic quantities, particularly the coercive field ($H_C$) and remanence ($M_R$) obtained from magnetization–field [$M(H)$] measurements. Well defined values can be obtained both for uniaxial- and cubic-anisotropy particles \cite{garcia1999monte,garcia1999influence}. Thus, experimental $M(H)$ data can be used to infer the underlying anisotropy characteristics \cite{wernsdorfer2002magnetisation,peterson2021determination}.
Owing to its simplicity and physical transparency, the SW framework has been extensively applied in both experimental \cite{cimpoesu2013generalized,winter2010interpretation} and theoretical \cite{yan2013modification,amanoloaei2021dynamic,lee2024statistical} studies to interpret the magnetization behavior of MNP ensembles.

The objective of this work is to investigate the applicability of an extended SW model to account for actual particle shape of experimental samples. Thus, while Stoner and Wohlfarth considered particles with either uniaxial ($K_u$) or cubic ($K_c$) anisotropy constants in their pioneering work \cite{Stoner1948} -i.e. one type or the other-, recent works identified the need to account for both anisotropies simultaneously: the intrinsic magnetocrystalline term, always present, plus a uniaxial contribution accounting for shape effects \cite{coffey2020thermal,garanin2003surface,kachkachi2007effects,gandia2020elucidating,russier2016blocking}. From now on, we shall refer to such extended SW model as the $K_c+K_u$ model.

\begin{figure}[!thp]
\centering
\includegraphics[width=0.9\columnwidth]{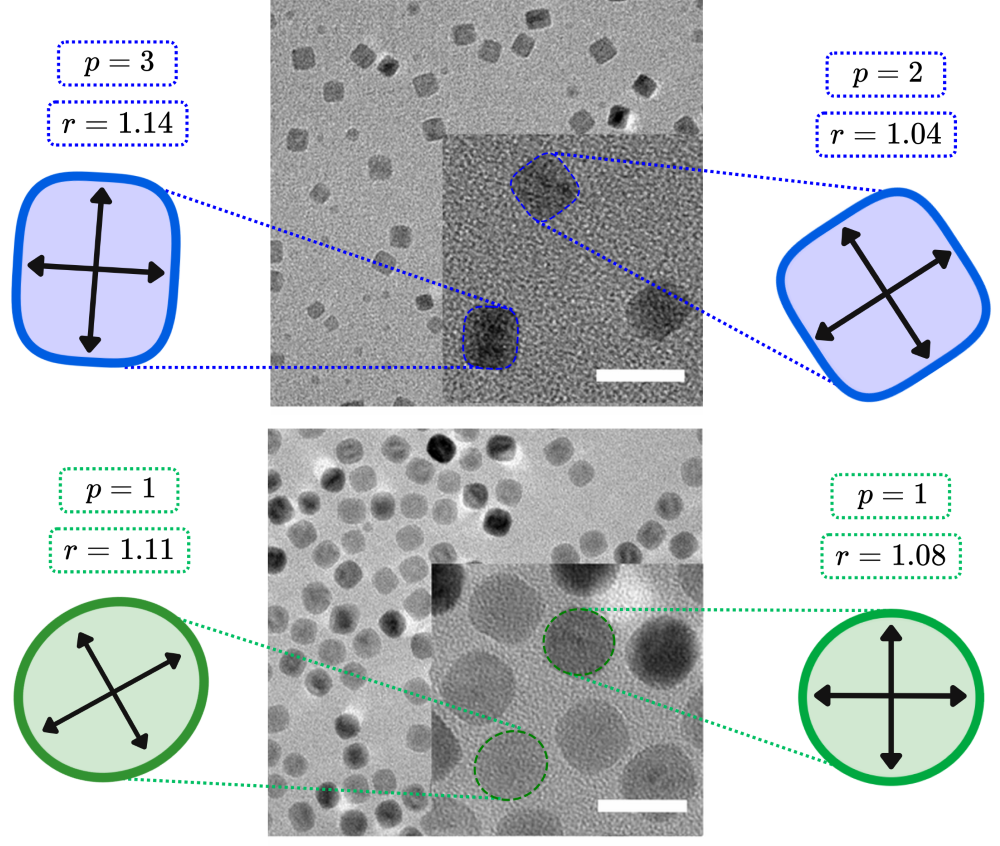}
\caption{\label{Fig_xperim_examples} Schematic representation of superellipsoidal shapes of different exponents $p$ and axial ratios $r = c/a$, %. It can be seen how small deviations from spherical or cubic symmetry give rise to more flattened or elongated shapes,
similar to those observed experimentally. TEM images modified
from \cite{paez2023optimization} with the permission of American Chemical Society.}
\end{figure}

Despite the reported works based on the more sophisticated $K_c+K_u$, the fact is that the behavior of MNPs is frequently analyzed within an effective uniaxial-anisotropy framework \cite{conde2015single,niculaes2017asymmetric}, assuming that the shape-induced anisotropy contribution largely dominates over the intrinsic magnetocrystalline one \cite{usov2010low,vallejo2013effect}. In other words, it is implicitly presumed that the particles deviate substantially from a spherical geometry. In practice, this assumption is not unfounded, as MNPs are rarely perfect spheres or cubes \cite{salazar2008cubic, paez2023optimization, roca2019design, navarro2019slow, gavilan2023scale, mandriota2025magnetic, muro2019precise}, as illustrated in Fig.~\ref{Fig_xperim_examples}. The extent to which deviations from ideal geometry modify the overall magnetic response remains an open question.%, which the present study seeks to address. ESTO É O QUE DICIMOS NA 1ª FRASE DO PARÁGRADO SEGUINTE...

%However, while such an \textit{effective} uniaxial-anisotropy approximation may be reasonable at high fields, we have recently shown \cite{Failde2024} that it can lead to significant deviations at moderate field amplitudes, such as those relevant to magnetic hyperthermia.

The present study aims to establish the range of validity of the simplified uniaxial-only anisotropy approach in comparison with the more sophisticated $K_c+K_u$ model. To this end, we have conducted a computational simulation of $M(H)$ hysteresis loops of particles with different shapes and sizes, considering both a full description of the particle properties and the corresponding macrospin approximation. 
A micromagnetic approach is employed to accurately capture particle shape while maintaining computational feasibility for relatively large system sizes.
The size dependence is analyzed to assess the applicability of the macrospin approximation. For simplicity, we restrict the analysis to spherical and cubic particles, as well as intermediate shapes described by a superellipsoidal geometry, including elongation effects.

The article is organized as follows. Sections~\ref{sec_physical_model} and~\ref{Sec_Computational} briefly describe the physical and computational frameworks employed. In Section~\ref{role}, we analyze in detail the influence of particle shape and elongation on the magnetic behavior of magnetite nanoparticles from a micromagnetic perspective, focusing on the determination of the size range for which %the critical volume $V_c$ below which 
the magnetization remains essentially uniform under an applied external field. In Section~\ref{Macro_vs_micro_sec}, we assess the validity of the generalized $K_c + K_u$ macrospin model%, which incorporates both cubic and effective uniaxial anisotropy contributions,
through a direct comparison with micromagnetic simulations. This analysis includes the angular dependence of the main hysteresis parameters, orientation-averaged hysteresis loops, and a phase diagram identifying the range of applicability of the macrospin approximation. Finally, in Section~\ref{SW_model}, we contrast these results with those obtained from the SW model considering only uniaxial or cubic anisotropy.% highlighting its limitations when applied to iron oxide nanoparticles (IONPs).

%\CommOscar{ATA AQUÍ DE MOMENTO... quizais se poida usar o que fixeches antes, Iago. Pero necesito desconectar un rato agora... a ver qué vos parece de momento o cambio de enfoque.} 

%\CommIago{Me gusta la estructura de introducir el SW, hablar de las limitaciones y las goemetrías experimentales, y luego decir que vamos a ver el papel de estas formas y su descripción mediante modelos más sencillos. Lo único que me queda un poco extenso. Seguro que se puede hilar mejor y más breve.}

\section{Physical model}\label{sec_physical_model}

In this work, we consider two complementary descriptions of the magnetic behavior of MNPs. On the one hand, we employ a full micromagnetic model, in which the particle shape is explicitly accounted for through a superellipsoidal parametrization. On the other hand, we use a macrospin model, where the particle is assumed to remain in a quasi-uniform magnetic state and the effect of elongation is incorporated effectively through the demagnetizing factors entering the energy terms. The material parameters used %throughout the study 
are listed in Table~\ref{tab:parameters}.% and correspond to representative bulk values reported in the literature.

\begin{table}[H]
\centering
\caption{Material parameters used for magnetite (Fe$_3$O$_4$).}
%\caption{Material parameters used for magnetite(\texorpdfstring{Fe$_3$O$_4$}{Fe3O4}).}
\label{tab:parameters}
\begin{ruledtabular}
%\begin{tabular}{@{}p{4cm}p{1.5cm}p{2.5cm}@{}}
\begin{tabular}{lll}
%\hline
\textbf{Parameter} & \textbf{Symbol} & \textbf{Value} \\ %\hline
Saturation magnetization & $M_s$ & $4.8\times10^5$~A/m \\
Exchange stiffness constant & $A$ & $1.1\times10^{-11}$~J/m \\
Cubic anisotropy constant & $K_c$ & $-1.1\times10^{4}$~J/m$^3$ \\
Exchange length & $L_{\mathrm{ex}}$ & $8.7$~nm %\\ \hline
\end{tabular}
\end{ruledtabular}
\end{table}

\subsection{Micromagnetic model}

\begin{figure}[!thp]
\centering
\includegraphics[width=0.95\columnwidth]{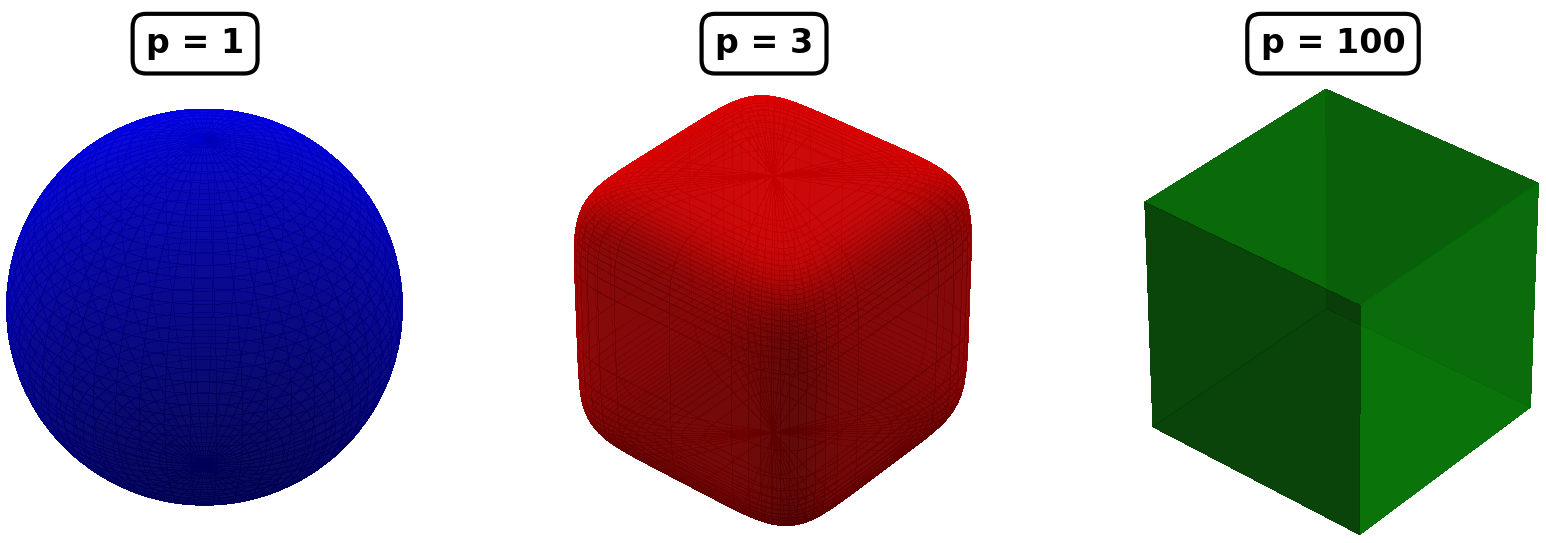}
\caption{\label{Fig_scheme-shapes}Schemes of different example geometries studied in this work: $p= 1$ (sphere),  $p= 3$, and $p= 100$ (cube), for the particular case of $r=1.0$}
%\label{Fig_scheme-shapes}
\end{figure}

The physical model employed in the micromagnetic simulations is the same as that used in Ref.~\cite{Lopez-Vazquez_JMMM_2026}. Briefly, the micromagnetic model accounts for four energy contributions: cubic magnetocrystalline anisotropy, exchange, demagnetizing, and Zeeman energies. We consider nanoparticles exhibiting variations in both shape and elongation, modelled through the geometry of a generalized superellipsoid \cite{Jaklic2000}. This parametrization enables a continuous description of morphologies ranging from spherical to cubic, as well as deviations from perfect isotropy through an axial ratio $r \neq 1.0$, defined as the ratio between the longest and shortest particle axes ($r = c/a$). Such variations provide a realistic representation of experimentally synthesized magnetite nanoparticles, which almost invariably display slight elongations and departures from ideal shapes, as shown in Fig.~\ref{Fig_xperim_examples}. The superellipsoidal geometry is defined by \cite{donaldson2017nanoparticle,Dresen2021neither}
\begin{equation}
\left( \frac{x}{a} \right)^{2p} + 
\left( \frac{y}{b} \right)^{2p} + 
\left( \frac{z}{c} \right)^{2p} \leq 1,
\end{equation}
where $a, b, c$ are the particle semi-axes and $p$ is the shape exponent. In this study, we focus on three representative morphologies by varying the $p$ value: $p=1$, $p=3$, and $p=100$, as illustrated in Fig.~\ref{Fig_scheme-shapes}.  

The corresponding superellipsoidal volume (V) is calculated from the analytical expression 
\begin{multline}
   \frac{V}{abc}
		=
		2\varepsilon_1\varepsilon_2 
		B\left(\frac{\varepsilon_1}{2},\varepsilon_1+1\right)
		B\left(\frac{\varepsilon_2}{2},\frac{\varepsilon_2+2}{2}\right)\\
		= \frac{2}{p^2}
		B\left(\frac{1}{2p},\frac{2p+1}{2p}\right)
		B\left(\frac{1}{2p},\frac{p+1}{p}\right)\ ,
\label{Eq_VolSuperball_2}
\end{multline}
where \(\varepsilon_1\) and \(\varepsilon_2\) are the exponents of the general superellipsoidal parametrization, which in the present case reduce to \(\varepsilon_1=\varepsilon_2=1/p\), and \(B(x,y)=\Gamma(x)\Gamma(y)/\Gamma(x+y)\) is the beta function \cite{jiao2009optimal,Jaklic2000}. Throughout the first part of the work, the nanoparticle size is expressed as $V^{1/3}$. This normalization ensures that different geometries can be directly compared at equal volume, so that the observed differences in behavior arise solely from the particle shape and elongation.

\subsection{Macrospin model}

As introduced above, we also consider a simplified macrospin description, denoted as the $K_c + K_u$ model, in which the nanoparticle is assumed to remain in a quasi-uniform magnetic state. This model combines two anisotropy contributions: (i) the cubic magnetocrystalline anisotropy intrinsic to magnetite, and (ii) an effective uniaxial shape anisotropy associated with particle elongation. In the presence of an applied field, the total energy density ($\mathrm{J/m^3}$) of the model is given by the sum of the cubic, uniaxial, and Zeeman terms:
\begin{multline} \label{energy}
E = K_{c}\,(\alpha_1^2\alpha_2^2 + \alpha_2^2\alpha_3^2 + \alpha_3^2\alpha_1^2) 
    + K_{u}\sin^2\theta \\
    - \mu_0 M_s H \cos(\theta_H - \theta),
\end{multline}
where \(K_c\) and \(K_{u}\) are the cubic and uniaxial shape anisotropy constants, respectively; \(\alpha_i\) (\(i=1,2,3\)) are the direction cosines of the magnetization relative to the crystallographic axes; \(\theta\) and \(\theta_H\) are the angles that the easy axes forms with magnetization and applied field directions, respectively.% (taken as the \(z\) axis); and \(\theta_H\) denotes the angle between the applied field and the same axis.

The uniaxial shape anisotropy constant is obtained as
\begin{equation}
K_{u} = \frac{\mu_0}{2}\,(N_a - N_c)\,M_s^2,
\label{K_sh}
\end{equation}
where $N_a$ and $N_c$ are the demagnetizing factors corresponding to the minor and major axes of an ellipsoid, respectively \cite{cullity2011introduction}. These factors depend only on the axial ratio $r = c/a$ of the nanoparticle (with the long axis oriented along the $z$ direction) and are expressed as
\begin{equation}
N_c(r) = \frac{1}{r^2 - 1}\left[\frac{r}{\sqrt{r^2 - 1}}
\ln\left(r + \sqrt{r^2 - 1}\right) - 1\right],
\end{equation}
\begin{equation}
N_a(r) = \frac{1 - N_c(r)}{2}.
\end{equation}

\begin{figure}[!thp]
    \centering
    \includegraphics[width=0.85\columnwidth]{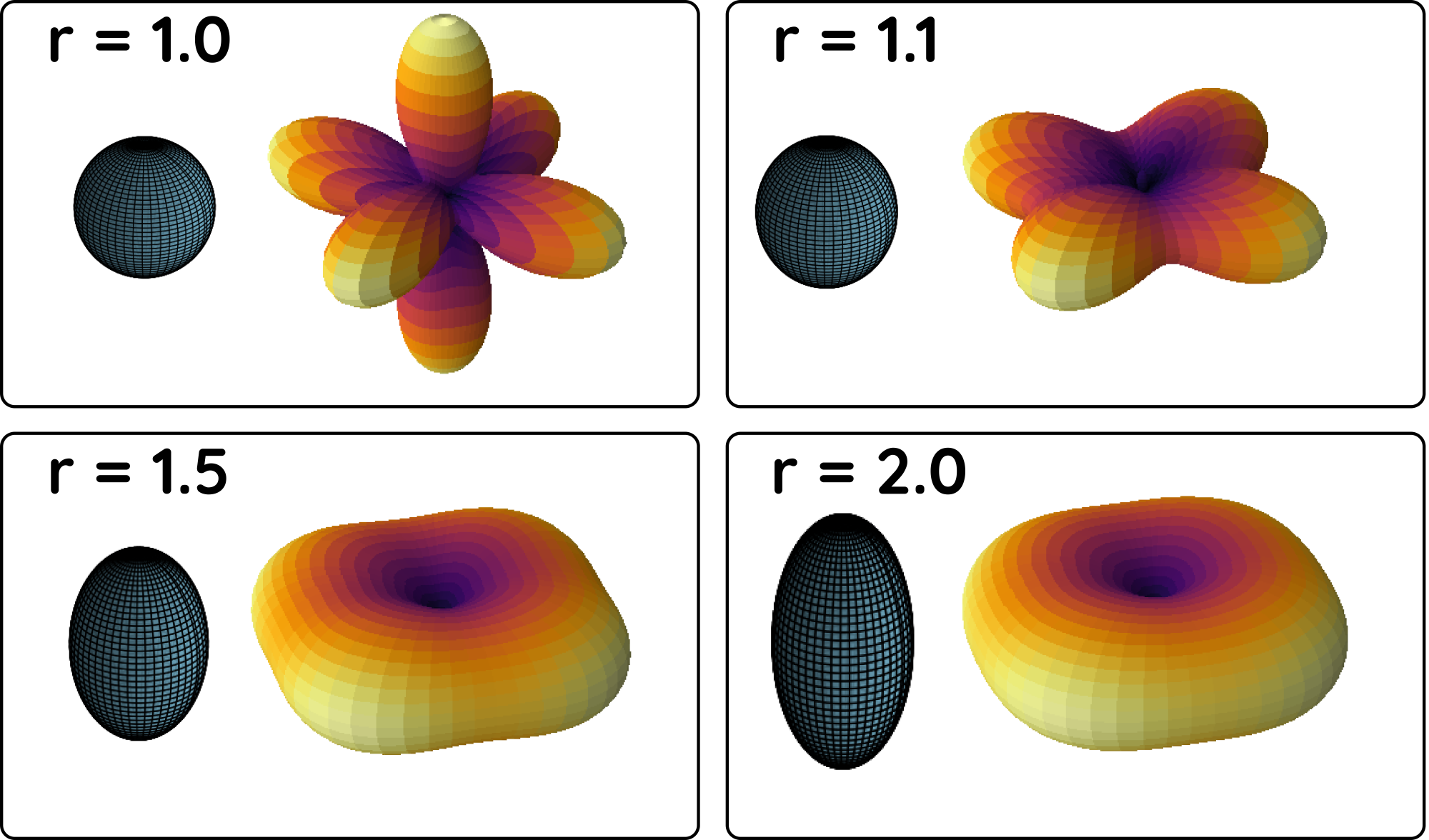}
    \caption{Schematics of particles with different elongations (in grey), and corresponding energy landscapes.% in spherical coordinates (right) and corresponding nanoparticle geometries (left) for different axial ratios \(r\).
    }
    \label{fig:barreras}
\end{figure}

The combined effect of the cubic and uniaxial anisotropy terms determines the energy landscape of the particle in absence of an external magnetic field. 
Fig.~\ref{fig:barreras} shows these energy surfaces in spherical coordinates for different axial ratios $r$, for several particle elongations.

\section{Computational details}\label{Sec_Computational}

To carry out the simulations, we employ the Object Oriented MicroMagnetic Framework (\textit{OOMMF}) software~\cite{oommf}, a finite-difference solver widely used for the study of static and dynamic magnetization processes in nanoscale systems. As discussed previously, we compare two complementary modelling approaches throughout this work: a full micromagnetic description and simplified macrospin models. The specific computational details of each approach are described below.

\begin{figure}[!thp]
\centering
\includegraphics[width=0.6\columnwidth]{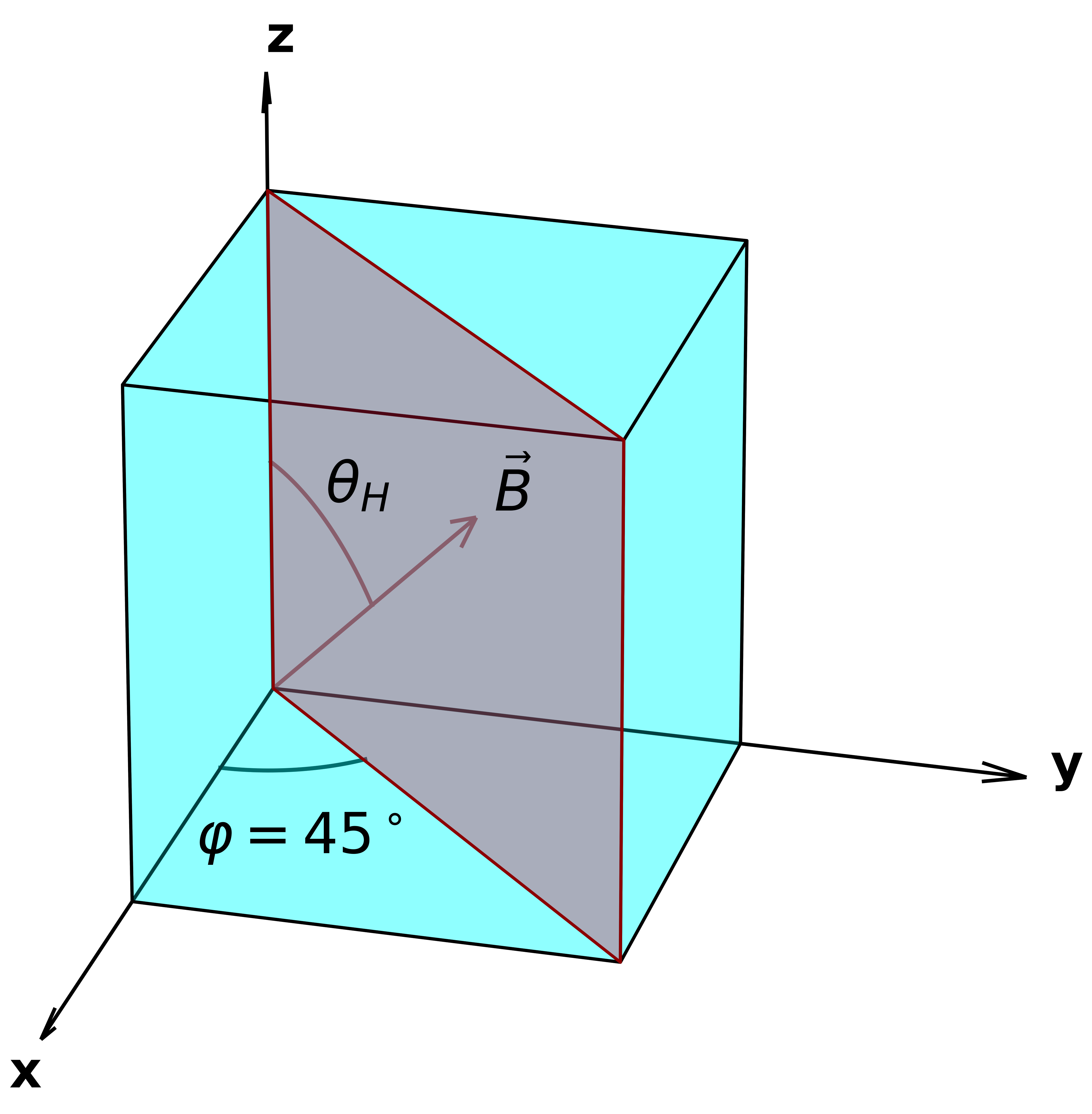}
\caption{Schematics of the applied field direction regarding an elongated cubic nanoparticle (\(p = 100\), \(r = 1.2\)). The field lies in the \(\varphi = 45^\circ\) plane, while the \(\theta_H\) value is varied.}
\label{field-directions}
\end{figure}

Since our aim here is to investigate how particle shape and elongation influence the magnetic response of MNPs under an applied field, we focus on the calculation of magnetic hysteresis loops. Considering that the simulations are performed for individual nanoparticles, the response of a disordered ensemble must be approximated by considering different orientations of the applied field with respect to the particle anisotropy axes. In the ideal case, this would require averaging over infinitely many field directions. Since such a procedure is not computationally feasible, we restrict the analysis to a selected set of representative field orientations, schematically shown in Fig.~\ref{field-directions}. As indicated in the figure, the applied field is taken within the plane defined by $\varphi=45^\circ$, while the polar angle $\theta_H$ is varied. This choice is motivated by the cubic magnetocrystalline anisotropy of magnetite ($K_c<0$), for which the easy directions lie along the $\langle 111\rangle$ body diagonals of the cubic lattice. By selecting $\varphi=45^\circ$ and sweeping $\theta_H$, the applied-field direction is constrained to a plane that intersects these diagonal easy axes and, depending on particle elongation, also crosses the effective minimum-energy direction resulting from the competition between cubic and uniaxial anisotropies. The elongation-dependent equilibrium easy direction obtained from the combined $K_c+K_u$ energy is discussed in Appendix~\ref{minima}.

For each field value, the equilibrium magnetization configuration is obtained using the OOMMF \textit{Oxs\_MinDriver} energy minimization module, with a convergence criterion of \textit{stopping m$\times$H$\times$m} = 0.1. To ensure numerical stability and avoid trapping in local minima, simulations were repeated with different initial magnetization states (uniform, random, and vortex-like), consistently converging to the same equilibrium configuration.

\subsection{Micromagnetic simulations}

In the micromagnetic framework, each nanoparticle is discretized into cubic cells of 1~nm, a value smaller than the exchange length of magnetite (see Table. \ref{tab:parameters}), thus ensuring that the exchange interaction is properly resolved. To verify that this discretization does not bias the results, representative simulations were repeated using finer meshes, yielding 
equilibrium configurations within numerical accuracy. Since no appreciable changes were observed while the computational cost increased substantially, a cell size of 1~nm was adopted for all micromagnetic calculations reported here. For this case, the field direction is varied in steps of $5^\circ$ in $\theta_H$.

\subsection{Macrospin simulations}

In the macrospin approach, the nanoparticle is represented by a single uniform magnetization vector $\boldsymbol{\mu} = M_s V\,\mathbf{m}$, whose behavior is governed by the different energy contributions defined in the previous section (Eq.~\ref{energy}). Taking advantage of the much lower computational cost of this approach, the macrospin hysteresis loops are calculated using a finer angular resolution of $0.1^\circ$ in $\theta_H$.

\section{Results}

The results are presented in three stages. We first analyze, from a micromagnetic perspective, how particle shape and axial ratio influence the magnetic response of the nanoparticles under an applied field, with particular emphasis on identifying the size range over which the magnetization remains essentially quasi-uniform. Next, within this regime, we assess the validity of the $K_c+K_u$ macrospin model through direct comparison with the micromagnetic simulations. Finally, we compare this description with simpler SW-type models.% in order to highlight the limitations of these approaches relative to the combined model.

\subsection{Micromagnetic simulations}\label{role}

To assess the feasibility of using effective macrospin models to describe particles with different shapes and elongations, the first requirement is to verify that the magnetization behaves essentially uniform. That is to say, one must examine how particle size and shape affect the magnetization behavior. 

To this end, we build upon our previous work \cite{Lopez-Vazquez_JMMM_2026}, where we investigated the influence of particle shape, size, and elongation on the spontaneous magnetization of MNPs using the same superellipsoidal geometry. In that study, performed in the absence of an external applied magnetic field, we found that both shape and elongation play a decisive role in determining the magnetic behavior of the system, particularly around what we shall refer to as the critical volume ($V_c$)-below which the magnetization remains essentially uniform, preserving the macrospin-like character of the particle. 

In the present work, we extend our previous analysis to the case where an external magnetic field is applied. By studying hysteresis loops for different geometries ($p = 1, 3, 100$) and elongations, we aim to evaluate how the field modifies the magnetic response and whether it affects the limits of validity of the quasi-uniform regime. Since this point is not the main focus of the present work, the corresponding analysis is provided in Appendix~\ref{critic}. As discussed there, when an external magnetic field is applied, the critical upper-size limits are found to lie in the range $V^{1/3} = 50 - 55 \, \mathrm{nm}$ for the three geometries considered and for elongations between $r = 1.0$ and $r = 1.2$. These values are comparable to those observed in the absence of a field \cite{Lopez-Vazquez_JMMM_2026}, indicating that the applied field does not significantly alter the upper limit of the quasi-uniform regime. Therefore, we can safely select particle sizes within this range to analyze in detail the influence of small variations in shape and size on magnetic behavior.

\begin{figure*}[!ptb]
\centering
\includegraphics[width=1.0\textwidth]{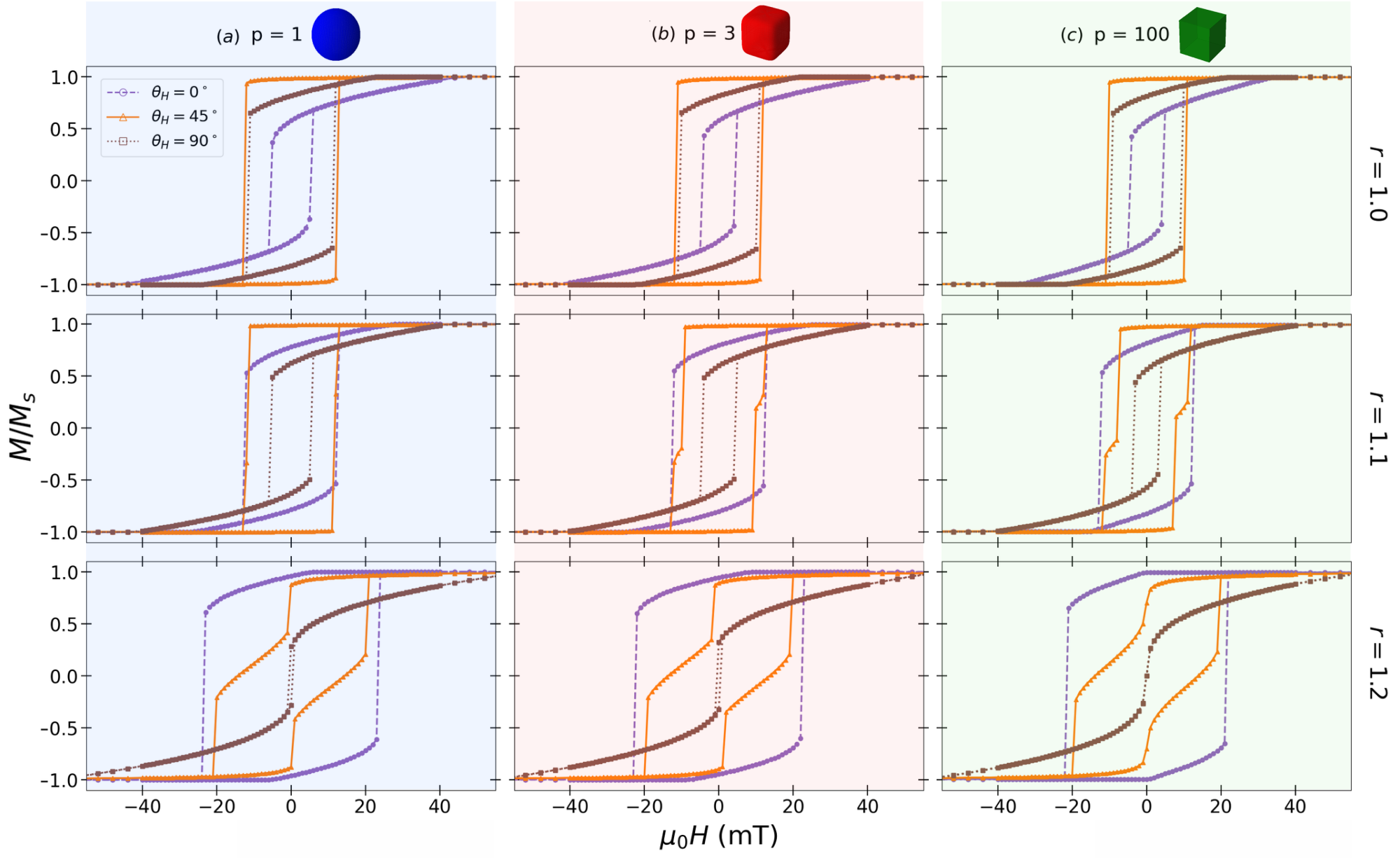}
\caption{%Angular-dependent 
Example hysteresis loops for a $V^{1/3}=40$ nm superellipsoidal MNP with different shape parameters (columns; $p=1,3,100$) and elongations (rows; $r=1.0,1.1,1.2$), %Columns and rows correspond to $p$ and $r$, respectively. The cases are 
for the field applied at representative directions $\theta_H=0^\circ$, $45^\circ$, and $90^\circ$.% with $V^{1/3}=40$ nm.
}
\label{elongation_examples}
\end{figure*}

To illustrate the variations in magnetic behavior arising from changes in particle shape and slight elongations, Fig.~\ref{elongation_examples} presents the hysteresis loops obtained for different directions of the applied magnetic field. Results are shown for the three geometries studied ($p = 1, 3, 100$) and three representative elongations ($r = 1.0$, $1.1$, and $1.2$), for particles with $V^{1/3} = 40$ nm, that is, within the range corresponding to essentially uniform behavior. For a more complete analysis including the full angular dependence, see Appendix~\ref{ap:4}.

%In Fig.~\ref{elongation_examples} it is observed that, in first approximation, variations in the particles axial ratio have a more significant influence on the magnetic response (more pronounced for higher $p$) than the particle shape in itself, which only in the transition from $p=1$ to $p=3$ shows a noticeable effect.%, affecting both the coercivity and the magnetization reversal process under the applied field. By contrast, somewhat unexpectedly, the particle shape does not appear to play a major role in the magnetic behaviour, since 
%Otherwise, the overall form of the hysteresis loops remains largely unchanged across the different values of the shape parameter $p$.

In Fig.~\ref{elongation_examples} it is observed that, in first approximation, variations in the axial ratio of the particles have a more significant influence on the magnetic response (more pronounced for higher $p$) than the particle shape in itself, which only in the transition from $p=1$ to $p=3$ shows a noticeable effect; otherwise, the overall form of the hysteresis loops remains largely unchanged across the different values of the shape parameter $p$. In other words, within the range of morphologies considered here, the magnetic response is essentially similar regardless of whether the particle geometry is closer to a sphere or to a cube. This suggests that particle elongation is the dominant factor for the magnetic behavior, with the shape playing a second-order correction. For a more detailed analysis, see Appendix~\ref{ap:4}.

\subsection{$K_c + K_u$ model vs. Micromagnetics}
\label{Macro_vs_micro_sec}

In this section, we assess the validity of the effective $K_c + K_u$ macrospin description by directly comparing its predictions with full micromagnetic simulations. The comparison is carried out in three steps. First, we analyze the angular dependence of the main hysteresis parameters for representative quasi-uniform particles. Next, we compare orientation-averaged hysteresis loops, which provide a more realistic description of particle ensembles. Finally, we extend the analysis to the full size--elongation parameter space in order to identify the range of applicability of the macrospin approximation.

\subsubsection{Comparison of angular dependence}

\begin{figure}[!thp]
\centering
\includegraphics[width=0.85\columnwidth]{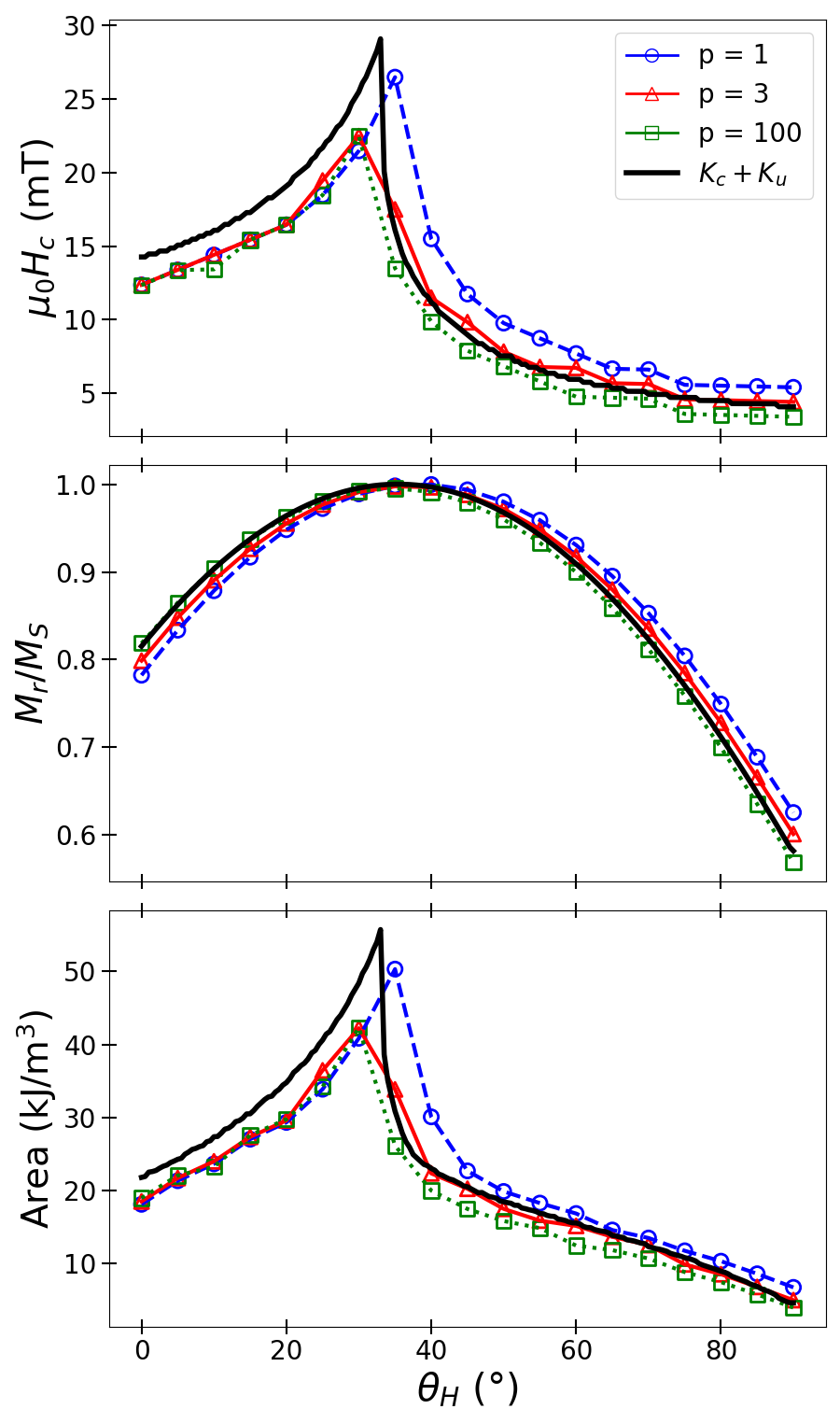}
\caption{Comparison of hysteresis parameters ($H_c$, $M_r$, and loop area) obtained from micromagnetic simulations (symbols) and the $K_c+K_u$ model (solid lines) for particles with $p=1,3,$ and $100$, $r=1.1$, and $V^{1/3}=40$.}

\label{macro_vs_micro}
\end{figure}

As a representative case, we focus on particles with \(V^{1/3} = 40\) nm and $r = $1.1, well below the critical threshold, and compare the resulting magnetic parameters ($H_c$, $M_r$, and hysteresis loop area) obtained from micromagnetic simulations to those obtained from the $K_c + K_u$ model. The results are shown in Fig.~\ref{macro_vs_micro}.

The results plotted in Fig.~\ref{macro_vs_micro} indicate an excellent agreement between the micromagnetic simulations and the $K_c + K_u$ model predictions for the ellipsoidal case (\(p = 1\)), as both models reproduce the same angular dependence of $H_c, M_r$, and hysteresis loop area. 
The small discrepancies observed between the two models arise from the finite discretization used in the micromagnetic calculations: with a spatial discretization of $1$ nm, it is not always possible to construct particles with exactly the same volume and axial ratio $r=1.10$. In practice, $r$ varies slightly between $1.08$ and $1.12$, which induces minor changes in the effective anisotropy field with respect to the idealized $K_c + K_u$ case.

The results for the superellipsoidal geometries (\(p = 3\) and \(p = 100\)) are also in very good agreement with the \(K_c + K_u\) model predictions, although the \(H_c\) values below \(35^\circ\) are slightly lower than those predicted by the macrospin model, again due to the discretization effects mentioned above. For a detailed analysis of the $H_C$ \textit{vs.} $\theta_H$ features the reader is referred to Appendix \ref{minima}.

Overall, these results show that the \(K_c + K_u\) approach remains valid across the full range of shapes considered here, from nearly spherical ($p = 1$) to nearly cubic ($p = 100$) particles, reaffirming that particle shape does not play a critical role in the magnetic behavior within the quasi-uniform regime. Although we only show the case \(r = 1.1\) here, the cases \(r = 1.0\) and \(r = 1.2\) exhibit the same good correspondence between micromagnetic simulations and the macrospin model, as will be shown more generally in the following subsection.

% \CommOscar{Respecto a la Fig. 6: mostramos r= 1.1 pero me pregunto si tenemos la misma para r= 1.2. Seguro que las variaciones en este caso son distintas (ya se ve en la Fig. 4). No sé si valdría la pena comentar. todo esto tiene relación con la Fig 12 del apéndice y la explicación que comento que deberíamos dar allí.}

% \CommIago{Acabo de hacer la figura, está en la carpeta figures como caso r = 1.2, por si la quieres comentar.} 
% \CommOscar{Vista!, el cambio es sustancial, para r= 1.2 ya domina la uniaxial y las barreras y máximo en Hc también}

\subsubsection{Comparison of orientation-averaged hysteresis loops}

To perform a more meaningful comparison, we now compare the averaged hysteresis loops obtained from micromagnetic simulations with those predicted by the effective \(K_c + K_u\) macrospin model. These averaged loops provide a more representative basis for comparison, as they reflect the expected magnetic behavior of an ensemble of MNPs with a quasi-random distribution of anisotropy-axis orientations. %Such a comparison allows us to evaluate the extent to which the $K_c + K_u$ model can describe experimentally relevant systems with dispersion in easy-axis directions. ACABAMOS DE DICILO...? REDUNDANTE??

\begin{figure}[!thp]
\centering
\includegraphics[width=0.85\columnwidth]{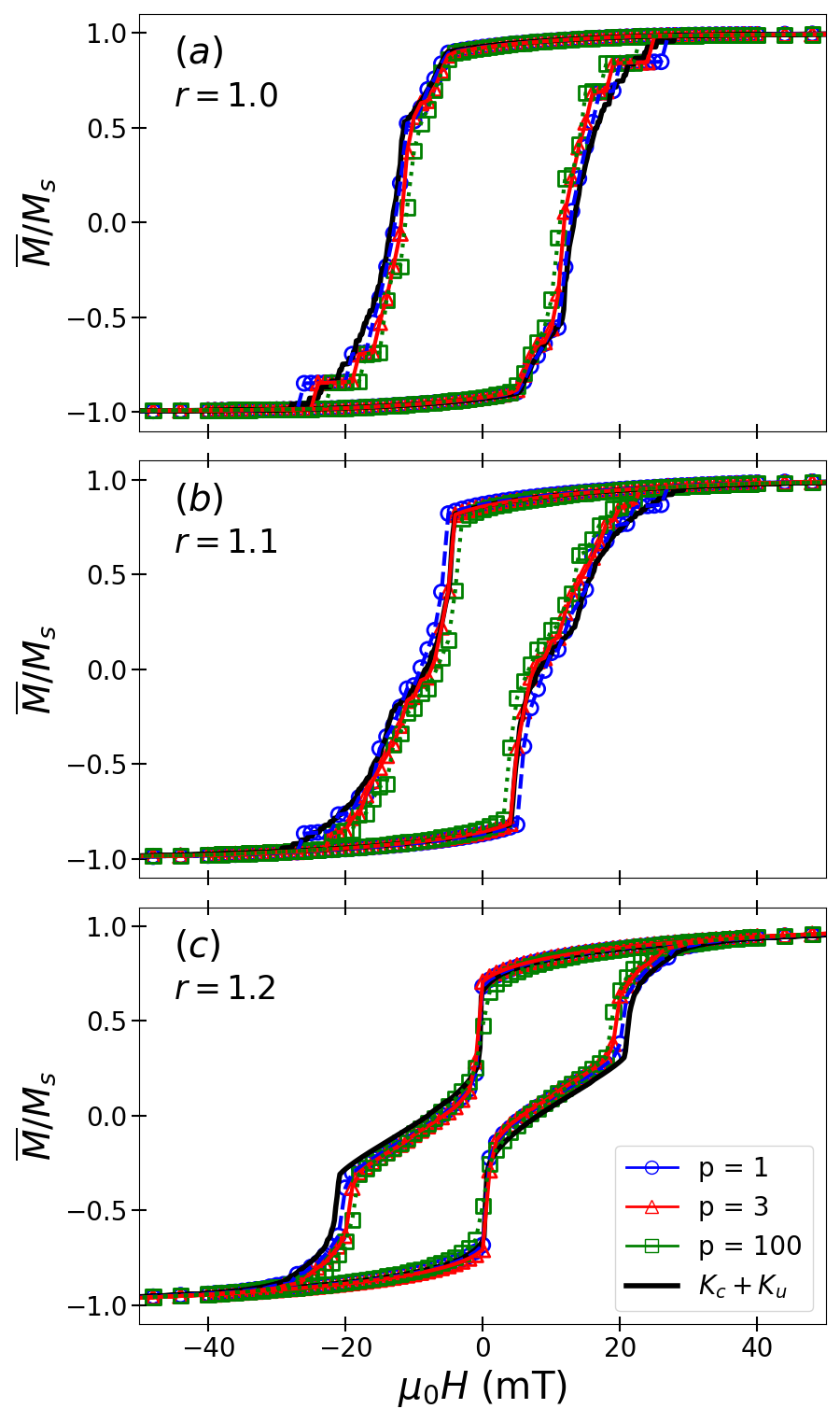}
\caption{Orientation-averaged hysteresis loops comparing micromagnetic simulations and the $K_c+K_u$ model for particles with $V^{1/3}=40$ nm and axial ratios $r=1.0$, $1.1$, and $1.2$.}
\label{macrospin_r_variable}
\end{figure}

The normalized averaged magnetization $\overline{M}(H)$, as a function of the applied field $H$, is computed as:
\begin{equation}
\overline{M}(H) = \frac{\int_0^{\pi/2} M(H, \theta_H) \cdot \sin(\theta_H) \, d\theta_H}{\int_0^{\pi/2} \sin(\theta_H) \, d\theta_H}  
\label{PROMEDIO}
\end{equation}
It is worth noting that this expression does not correspond to a fully isotropic (spherical) distribution, which would be computationally prohibitive, as explained in Sec.~\ref{Sec_Computational}. Instead, we consider a uniform distribution over the polar angle $\theta_H$, while fixing the azimuthal angle at $\varphi = 45^\circ$. This simplification provides a reasonable representative approximation to the angular averaging of the hysteresis response. The results are shown in Fig.~\ref{macrospin_r_variable}.

In Fig.~\ref{macrospin_r_variable} it is observed that the axial ratio has a pronounced influence on the hysteresis loop shape, notably affecting both coercivity and remanence. Despite these variations, the agreement between micromagnetic simulations and $K_c + K_u$ model remains excellent for all cases considered. This result confirms that the proposed framework accurately captures the magnetic response of superellipsoidal nanoparticle ensembles, even in the presence of moderate elongations or departures from perfect symmetry.

\subsubsection{Generalization: Phase Diagram} \label{Sec_Phase-Diagram}
Up to this point, our analysis has been limited to a narrow range of particle sizes and axial ratios, leaving open the question of whether the correspondence between micromagnetic simulations and the effective macrospin description persists across the full parameter space. The aim of this section is therefore to assess the validity range of the $K_c + K_u$ model and to identify the regions where ellipsoidal MNPs can still be accurately described within the macrospin approximation.

Based on the results of Sec.~\ref{Macro_vs_micro_sec}, where a nearly identical behavior was observed for the different geometries considered, we restrict the following analysis to ellipsoidal particles ($p=1$), assuming that these results are representative of other superellipsoidal shapes as well.

To explore the full parameter space, it is not feasible to compare the complete hysteresis loops for all particle sizes and axial ratios considered. It is therefore necessary to restrict the analysis to a representative parameter. For this purpose, we focus on the coercive field, which provides a sensitive measure of the deviations between the micromagnetic simulations and the $K_c + K_u$ model. This choice is further justified by the fact that other quantities, such as the remanence, exhibit only minor variations even when magnetization coherence is partially lost, whereas the coercive field is much more sensitive to the onset of non-uniform behavior (see Appendix~\ref{Mr} and Fig.~\ref{fig:diagrama_Mr}). The deviation is then evaluated using the mean absolute percentage error (MAPE) \cite{makridakis2008forecasting}:
\begin{equation}
\Delta H_c = 100 \times \frac{1}{n} \sum_{i=1}^{n} 
\left| \frac{H_{c,i}^{\text{micro}} - H_{c,i}^{\text{macro}}}{H_{c,i}^{\text{macro}}} \right|
\label{eq:mape}
\end{equation}
This expression allows us to quantify the discrepancy in relative terms with respect to the $K_c + K_u$ model prediction. However, it presents an important limitation: when the angle approaches $\theta_H = 90^\circ$, particularly for nanoparticles with large aspect ratios, the theoretical $H_c$ becomes very small. In this regime, even minor absolute differences can lead to large relative deviations, artificially amplifying the apparent error. To avoid this distortion, we excluded from the analysis the values corresponding to $\theta_H = 90^\circ$. 
% In addition, numerical convergence issues may occasionally arise for angles very close to $\theta = 0^\circ$, which also restricts direct comparison in those cases.

\begin{figure}[!thp]
    \centering
    \includegraphics[width=\columnwidth]{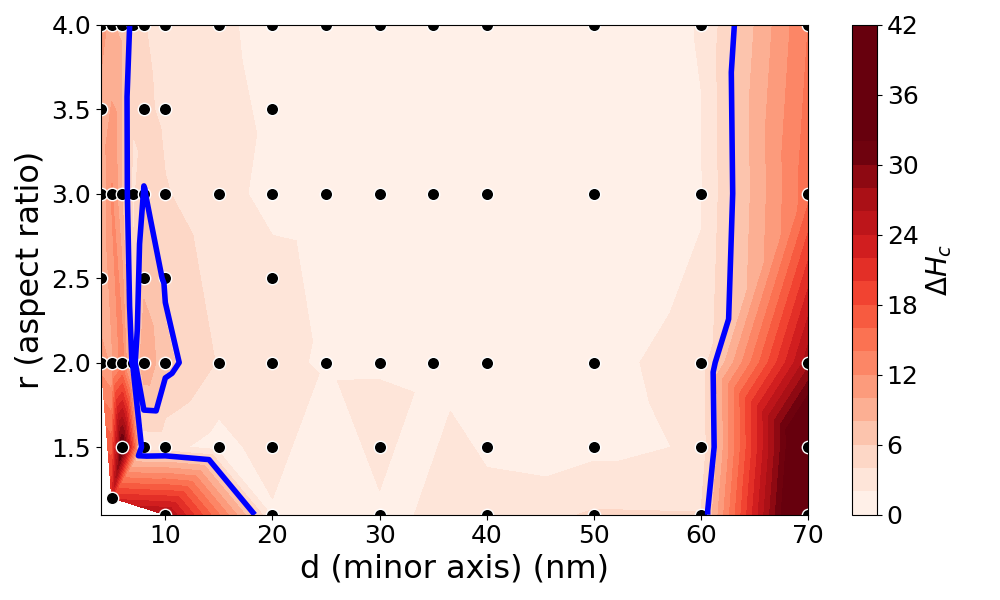}
    \caption{Mean deviation $\Delta H_c$ between coercive fields from micromagnetic simulations and the $K_c+K_u$ model as a function of $d$ and $r$. The blue contour identifies the region where deviations remain under $5\%$.}
    \label{fig:diagrama_completo-macrospin}
\end{figure}

Throughout this section, the particle size is expressed in terms of the minor-axis length, denoted by $d$. This choice is convenient because fixing $d$ and varying the aspect ratio $r$ enables a systematic construction of the phase diagram. In contrast, using the effective magnetic size $V^{1/3}$ would introduce an additional complication: due to the spatial discretization of $1$ nm$^3$ cells in the simulations, the represented volume changes discontinuously, making it impossible to maintain a constant $V^{1/3}$ when modifying $r$. The results are shown in Fig.~\ref{fig:diagrama_completo-macrospin}.

\begin{figure}[!thp]
    \centering
    \includegraphics[width=0.8\columnwidth]{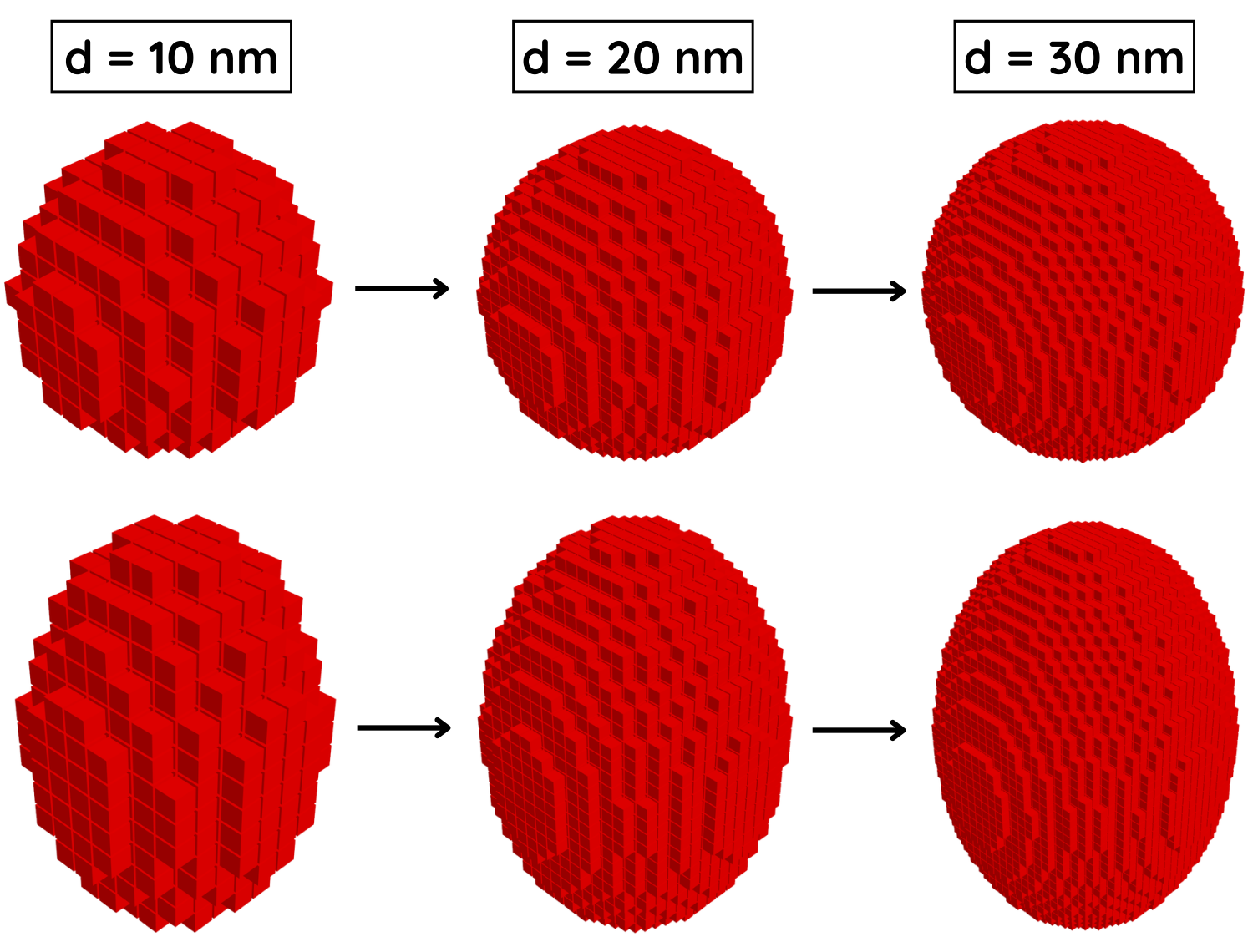}
    \caption{\label{fig:discretization} 
Discretization effects in spherical (top row) and elongated (bottom row) nanoparticles for different sizes $d=10, 20, 30$ nm.}
\end{figure}

The diagram displayed in Fig.~\ref{fig:diagrama_completo-macrospin} represents the mean deviation $\Delta H_c$ across the explored parameter space, where black dots indicate the specific $(d,r)$ combinations explicitly simulated. The color map is obtained by interpolation to guide the visualization of trends. The micromagnetic results deviate from the $K_c + K_u$ model predictions for small sizes, particularly below $d = 10$ nm, where the discrepancy exceeds 5$\%$ (as outlined by the blue contour). This deviation does not stem from a breakdown of the single-domain assumption, but rather from numerical artifacts associated with the finite spatial discretization. As illustrated in Fig.~\ref{fig:discretization}, when the particle size approaches $d = 10$ nm, the geometry becomes poorly resolved by the 1-nm cell size used in the mesh, leading to spurious variations in the computed coercivity. In this size regime, an atomistic description would be necessary to accurately capture the magnetic behavior.

At the opposite end of the size range, deviations above 5$\%$ are also observed for $d > 60$ nm.
In this case, the origin is physical: micromagnetic simulations indicate the onset of non-uniform magnetization states, implying that the particles no longer behave as coherent macrospins and cannot be accurately described within the $K_c + K_u$ framework.

\subsection{SW model vs. Micromagnetics}
\label{SW_model}

As demonstrated in the previous section, the $K_c + K_u$ model, combining cubic magnetocrystalline and uniaxial shape anisotropies, provides an excellent description of the micromagnetic behavior of nanoparticles below 60~nm, regardless of their specific geometry. 

In this section, we aim to place these results within the general context of the classical SW model, which, often formulated in terms of a uniaxial anisotropy, remains the most widely used theoretical approach in the literature for describing monodomain MNPs~\cite{stoner1948mechanism}. We therefore compare our micromagnetic results with the predictions of this classical SW framework in order to assess its suitability for describing the magnetic behavior of IONPs. For completeness, since magnetite IONPs intrinsically exhibit cubic magnetocrystalline anisotropy, we further include a comparison with a purely cubic variant of the SW model, in which the nanoparticle is assumed to be subjected exclusively to the cubic anisotropy contribution.

\subsubsection{Uniaxial SW model vs. Micromagnetics}
\begin{figure}[!thp]
    \centering
    \includegraphics[width=\columnwidth]{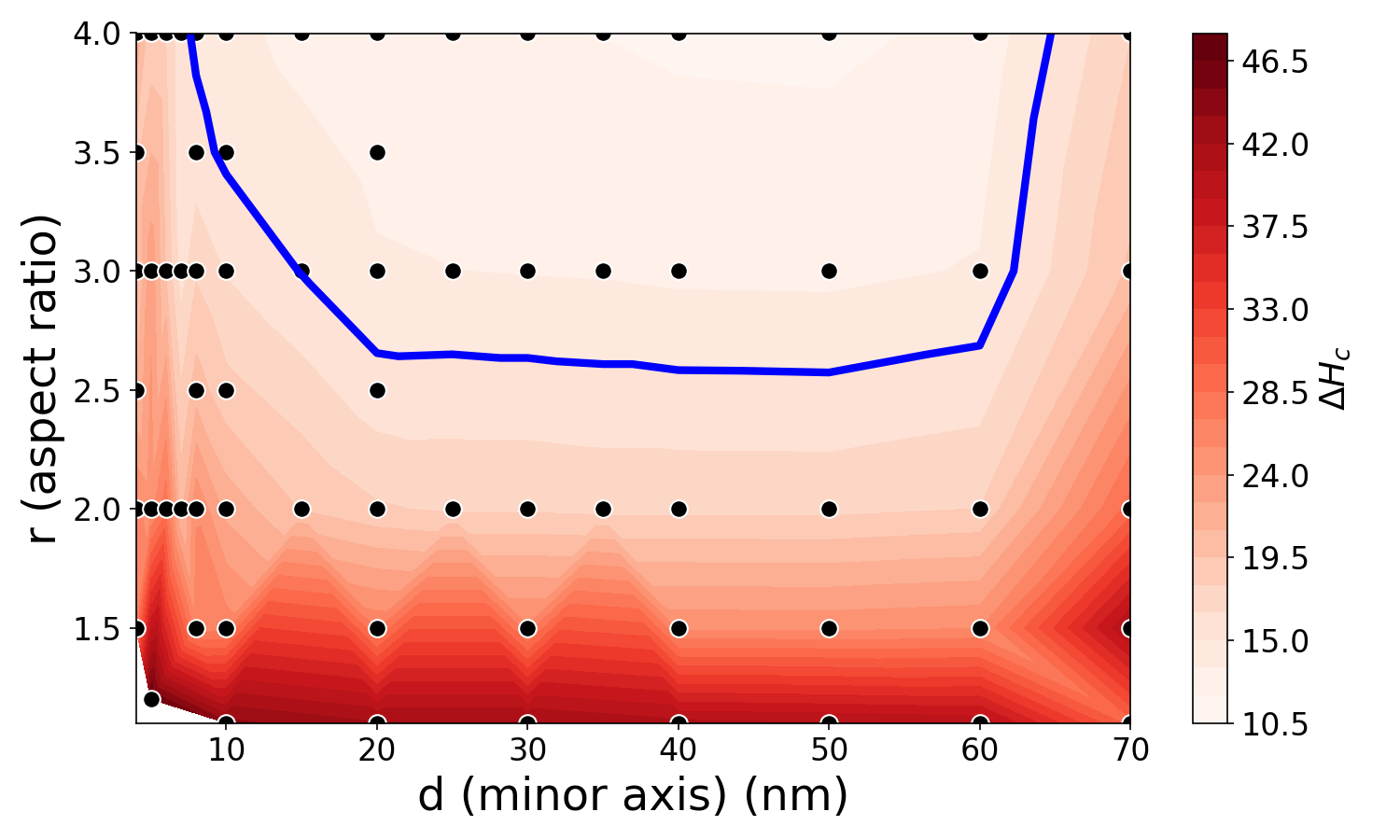}
    \caption{Mean deviation $\Delta H_c$ between coercive fields from micromagnetic simulations and the uniaxial SW model as a function of $d$ and $r$. The blue contour highlights the region with deviations below $15\%$, while all deviations are seen to remain above $10\%$.}
    \label{Fig_SW}
\end{figure}

We construct the same type of phase diagram introduced in Sec.~\ref{Sec_Phase-Diagram}, but taking now the uniaxial SW model as the theoretical reference. The resulting deviations are summarized in Fig.~\ref{Fig_SW}.
For all sizes and axial ratios considered, the deviations with respect to the SW model exceed 
$10 \%$.
This result underscores the non-negligible influence of cubic magnetocrystalline anisotropy on the magnetic behavior of magnetite nanoparticles, contrary to the common assumption that shape anisotropy alone suffices to describe their response \cite{usov2010low,vallejo2013effect}.
As in the case of the $K_c + K_u$ model comparison discussed in Sec.~\ref{Sec_Phase-Diagram}, deviations become more pronounced for the smallest particles (where numerical discretization effects become significant) and for the largest ones (where quasi-uniform magnetization is lost).
It is also worth noting that particles with nearly spherical shapes ($r\approx 1.0$) display particularly large discrepancies with respect to the uniaxial SW model. In these cases, the contribution from shape anisotropy weakens, thus allowing the cubic magnetocrystalline anisotropy to dominate the magnetic energy landscape and substantially affect the reversal process \cite{Failde2024}.

This behavior can also be interpreted in light of the anisotropy energy landscapes shown in Fig.~\ref{fig:barreras}.
For $r=1.0$, the anisotropy energy is purely cubic, giving easy axes along the crystallographic diagonals, as evidenced by the appearance of minima at intermediate directions. As the axial ratio increases, the contribution from uniaxial shape anisotropy becomes dominant, and the energy minima progressively align toward the elongated axis. Consequently, the system’s behavior approaches that predicted by the uniaxial SW model, explaining the gradual reduction of the coercive-field deviation to below 15\% in the region highlighted in Fig.~\ref{Fig_SW} with a blue contour.

\subsubsection{Cubic SW vs. Micromagnetics}

\begin{figure}[!thp]
    \centering
    \includegraphics[width=\columnwidth]{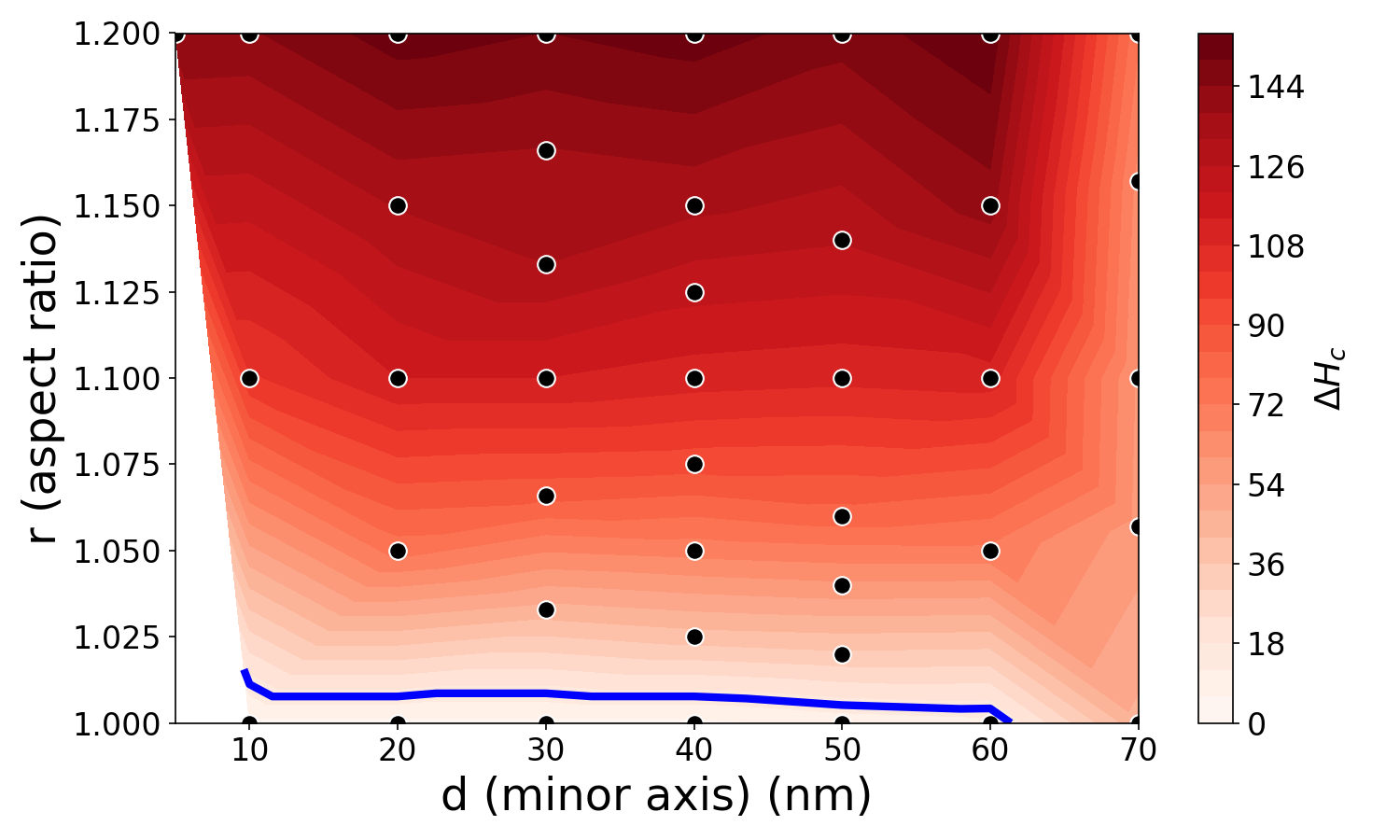}
    \caption{Mean deviation $\Delta H_c$ between coercive fields from micromagnetic simulations and the cubic SW model as a function of $d$ and $r$. The blue contour highlights the region with deviations below $15\%$.}
    \label{Fig_SW_Kc}
\end{figure}

After demonstrating the limited capability of the $K_u$-only SW model to describe the behavior of superellipsoidal MNPs, we now compare the micromagnetic results with the predictions of a purely cubic SW model, in which the nanoparticle is treated as a single-domain macrospin subject exclusively to cubic magnetocrystalline anisotropy. The resulting deviations between both approaches, computed according to Eq.~\ref{eq:mape}, are presented in Fig.~\ref{Fig_SW_Kc}.

The results shown in Fig.~\ref{Fig_SW_Kc} reveal that nanoparticles with axial ratios differing from \(r = 1.00\) display deviations exceeding 30\%, which increase further as elongation becomes more pronounced. Therefore, the cubic SW model is only capable of reproducing the magnetic behavior of nearly-spherical nanoparticles, for which uniaxial anisotropy is strictly absent. However, such ideal spherical systems are practically unattainable under experimental conditions \cite{salazar2008cubic, paez2023optimization}, strongly limiting the applicability of the $K_c$-only SW framework for realistic magnetite MNPs.

% \begin{figure*}[!ptb]
% \centering
% \includegraphics[width=1.0\textwidth]{Figures/Otra_posible.png}
% \caption{QUIZAS SE PUEDE COMBINAR DE ESTA FORMA Y COMENTAR ASI.}
% \label{otra_2}
% \end{figure*}

\section{Conclusions}

We have studied the feasibility of an extended Stoner--Wohlfarth model, denoted as the \(K_c + K_u\) model, to describe the magnetic behavior of magnetite nanoparticles with different shapes and axial ratios, defined through the superellipsoidal geometry. In this framework, the nanoparticle is treated as a single magnetic moment subject to two anisotropy contributions: the intrinsic cubic magnetocrystalline anisotropy of magnetite and an effective uniaxial term arising from particle elongation. In order to carry out a meaningful comparison between this reduced description and full micromagnetic simulations, it is first necessary to ensure that the nanoparticles remain within the quasi-uniform regime. For this reason, the first step of this study was to determine the critical size, which defines the boundary between quasi-uniform and non-uniform magnetic configurations, for our MNPs under the application of an external magnetic field. The results show that, for \(r = 1.1\), the critical upper-size limits for all geometries lie in the range \(50\text{--}55\) nm, independently of the precise particle shape, in good agreement with previous zero-field results reported in Ref.~\cite{Lopez-Vazquez_JMMM_2026}.

Within this essentially-uniform regime, our micromagnetic results show that the axial ratio plays the dominant role in the magnetic response, significantly affecting the coercivity, remanence, and magnetization reversal process under the applied field. By contrast, the particle shape itself, from nearly spherical to nearly cubic morphologies, does not appear to play a relevant role in the hysteresis behavior, acting at most as a second-order correction. In agreement with this, we observe a very good correspondence between micromagnetic simulations and the \(K_c + K_u\) model across the full range of superellipsoidal geometries considered here. This confirms that nanoparticles in the quasi-uniform regime can be accurately described within this reduced framework, regardless of their detailed morphology.

Given the widespread use of the uniaxial Stoner--Wohlfarth description in the literature for single-moment particle systems, we compared our micromagnetic results with two simplified SW models: a uniaxial-only case (\(K_u\)-only) and a cubic-only case (\(K_c\)-only). In both limits, deviations from the micromagnetic reference are substantial. The \(K_u\)-only model systematically fails to capture the intrinsic contribution of cubic magnetocrystalline anisotropy, even in particles where elongation is significant, whereas the \(K_c\)-only model is accurate only for nearly spherical particles and rapidly becomes inaccurate as the axial ratio increases. Thus, neither simplified SW description is sufficient to account for the magnetic behavior of realistic magnetite nanoparticles. By contrast, the \(K_c + K_u\) model provides a minimal yet quantitatively accurate description of the magnetic response within the quasi-uniform regime.

Beyond validating the $K_c + K_u$ description in the quasi-uniform regime, we establish a direct, quantitative link between particle geometry and the effective anisotropy parameters entering the macrospin model. This link provides a practical route for connecting micromagnetic simulations with analytical descriptions commonly used to interpret experimental data, particularly in realistic systems where nanoparticles rarely exhibit perfect shapes or strict monodispersity. The main contribution of the present work is therefore not simply the  combination of cubic and uniaxial anisotropies, which has already been explored in previous studies, including our own work\cite{Failde2024,usov2019heating,kalmykov2019forced,salvador2019competing}. Rather, the novelty is a systematic, quantitative correlation between particle shape and elongation and the effective parameters of the $K_c + K_u$ model. By showing that these parameters can be directly linked to realistic geometrical features of the nanoparticles, the present study moves beyond a purely phenomenological use of effective anisotropy terms and provides a physically motivated reduced description of magnetite MNPs within the quasi-uniform regime. More broadly, this geometry-anisotropy correspondence opens the way to the development of hybrid spin--particle dynamic models \cite{mamiya2011hyperthermic,mostarac2025thermal,wolfschwenger2024molecular,Usov2012,durhuus2024conservation,usadel2015dynamics,usadel2017dynamics,okada2025proposal}, in which geometrical constraints are expected to play a central role in determining the relative ordering and collective response of magnetic moments \cite{donaldson2017nanoparticle}.

The present study also suggests clear directions for future research. A first step would be a more detailed investigation of the small-size regime, where the applicability of the $K_c + K_u$ model becomes limited and an atomistic treatment is required \cite{moreno2020role}. Other extensions could include the incorporation of surface anisotropy \cite{garanin2003surface} and its interplay with finite-size effects \cite{kachkachi2000finite}, as well as the study of branch crossing \cite{mathews2020hysteresis} or doping effects \cite{omari2021effect}. Together, these efforts would further refine our understanding of the interplay between geometry, anisotropy, and magnetization dynamics in real nanoscale systems.

\begin{acknowledgments}
We acknowledge financial support by Spanish Ministerio de Ciencia, Innovaci\'on y Universidades through projects PID2019-109514RJ-100, PID2024-157172NB-I00 and CNS2024-154574, "ERDF A way of making Europe", by the "European Union", and Catalan DURSI (2021SGR0032). Xunta de Galicia is acknowledged for the projects ED431F 2022/005 and ED431B 2023/055. AEI is also acknowledged for  the \textit{Ram\'on y Cajal} grant RYC2020-029822-I that supports the work of D.S. We acknowledge the Centro de Supercomputacion de Galicia (CESGA) for computational resources.
\end{acknowledgments}

\appendix

\section{Behavior of the Coercivity Peak}
\label{minima}

In Sec. \ref{Macro_vs_micro_sec}, we reported the appearance of a pronounced peak in the coercive field as a function of the applied field angle $\theta_H$. We argued that this peak is associated with the direction of the easy magnetization axis. 
To substantiate this interpretation, we determine here the angle corresponding to the minimum of the anisotropy energy ($\theta_{\min}$), and compare it with the angles at which the coercivity reaches its maximum. For improved resolution, this comparison is performed using the results obtained from the combined $K_c + K_u$ anisotropy model.

We begin by considering the expression for the total anisotropy energy density of the system:
\begin{multline}
E_{\text{ani}}(\theta, \phi) = E_{\text{sh}}(\theta) + E_{\text{cub}}(\theta, \phi) \\
= K_{\text{u}} \sin^2\theta + K_c \left(\alpha_x^2 \alpha_y^2 + \alpha_y^2 \alpha_z^2 + \alpha_z^2 \alpha_x^2\right),
\end{multline}
where $\alpha_x = \sin\theta \cos\phi$, $\alpha_y = \sin\theta \sin\phi$, and $\alpha_z = \cos\theta$, and $K_{\text{u}}$ is obtained from Eq.~\ref{K_sh}. Simplifying for $\phi = 45^\circ$:
\begin{equation}
E_{\text{ani}}(\theta) = (K_{\text{u}} + K_c)\sin^2\theta - \frac{3}{4} K_c \sin^4\theta.
\end{equation}

Taking the derivative with respect to $\theta$, the condition for the minimum is:
\begin{equation}
\theta_{min} = \arcsin\!\left(\sqrt{\frac{2(K_{\text{u}} + K_c)}{3K_c}}\right).
\label{E_min}
\end{equation}

\begin{figure*}[!ptb]
\centering
\includegraphics[width=0.9\textwidth]{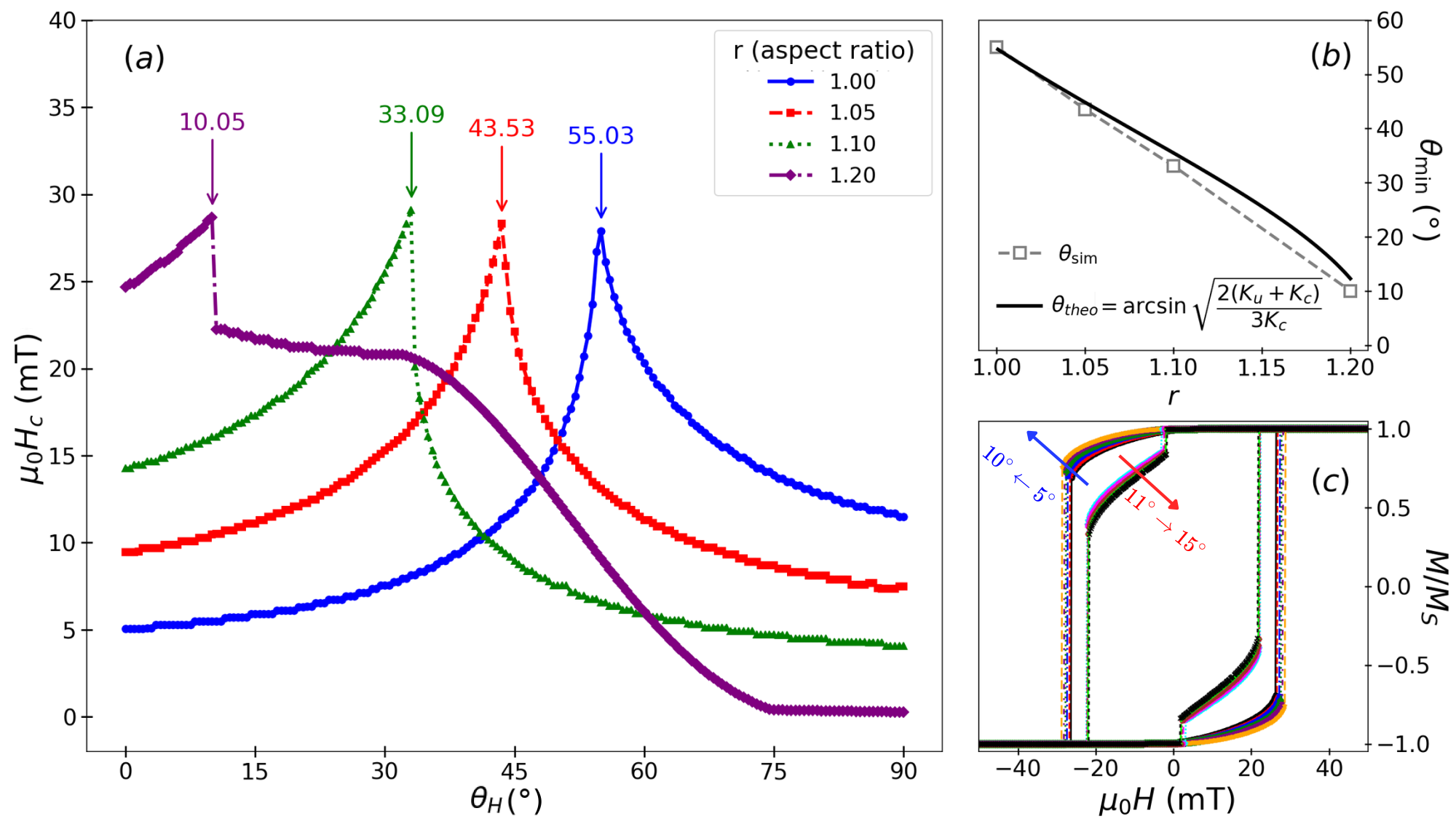}
\caption{(a) Dependence of the coercive field \(H_c\) on the field angle \(\theta_H\) for different axial ratios. The values where \(H_c\) is maximum are highlighted.  (b) Comparison between the theoretical and numerically obtained values of the angle corresponding to the minimum-energy direction, \(\theta_{\min}\), as a function of the axial ratio \(r\). (c) Hysteresis loops for the case \(r = 1.20\) in the angular region where the discontinuity is observed. All results shown were obtained from the \(K_c + K_u\) model.}
\label{fig:Hc_theta}
\end{figure*}

In Fig.~\ref{fig:Hc_theta}(a), we present the angular dependence of $H_c$ for different axial ratios, highlighting the emergence of a well-defined maximum. Fig.~\ref{fig:Hc_theta}(b) compares the angular positions of these $H_c$ peaks with the theoretical predictions for the energy minimum obtained from Eq.~\ref{E_min}. A very good agreement is observed between the angle of maximum $H_c$ and the direction of the easy magnetization axis. 
Furthermore, increasing the axial ratio leads to a systematic shift of the energy-minimizing angle, $\theta_{\min}$, toward the elongation direction of the particle. This behavior reflects the increasing dominance of shape anisotropy over magnetocrystalline anisotropy, as discussed in Sec.~\ref{role} (see, in particular, Fig.~\ref{fig:barreras}).
A notable change in behavior is observed for $r = 1.20$, where a discontinuity appears in the angular dependence of the coercivity at $\theta_{\min} = 10.05^\circ$. This discontinuity is associated with a qualitative change in the shape of the hysteresis loops, as illustrated in Fig.~\ref{fig:Hc_theta}(c). Remarkably, the angle at which this coercivity discontinuity occurs closely matches the theoretical prediction for the energy-minimizing direction, as shown in Fig.~\ref{fig:Hc_theta}(b).

\section{\label{critic} Quasi-Uniform Range}

To complement the discussion presented in Section~\ref{role}, here we provide the detailed analysis used to determine the critical sizes of micromagnetic MNPs under an applied magnetic field.

To perform a systematic study, we represent the characteristic parameters extracted from each hysteresis loop (namely $H_c$, $M_r$, and loop area) as a function of the field orientation angle $\theta_H$.
In this analysis, we focus on MNPs with an axial ratio of  $r = 1.1$, which corresponds to a typical elongation encountered in the synthesis of this type of nanoparticles \cite{salazar2008cubic}. Minor variations in $r$ between shapes arise from discretization effects.
We then investigate how the magnetic parameters of the hysteresis loops evolve with MNP size, with the goal of identifying signatures of the transition between quasi-uniform and nonuniform magnetization reversal modes.

\begin{figure*}[!ptb]
\centering
\includegraphics[width=1.0\textwidth]{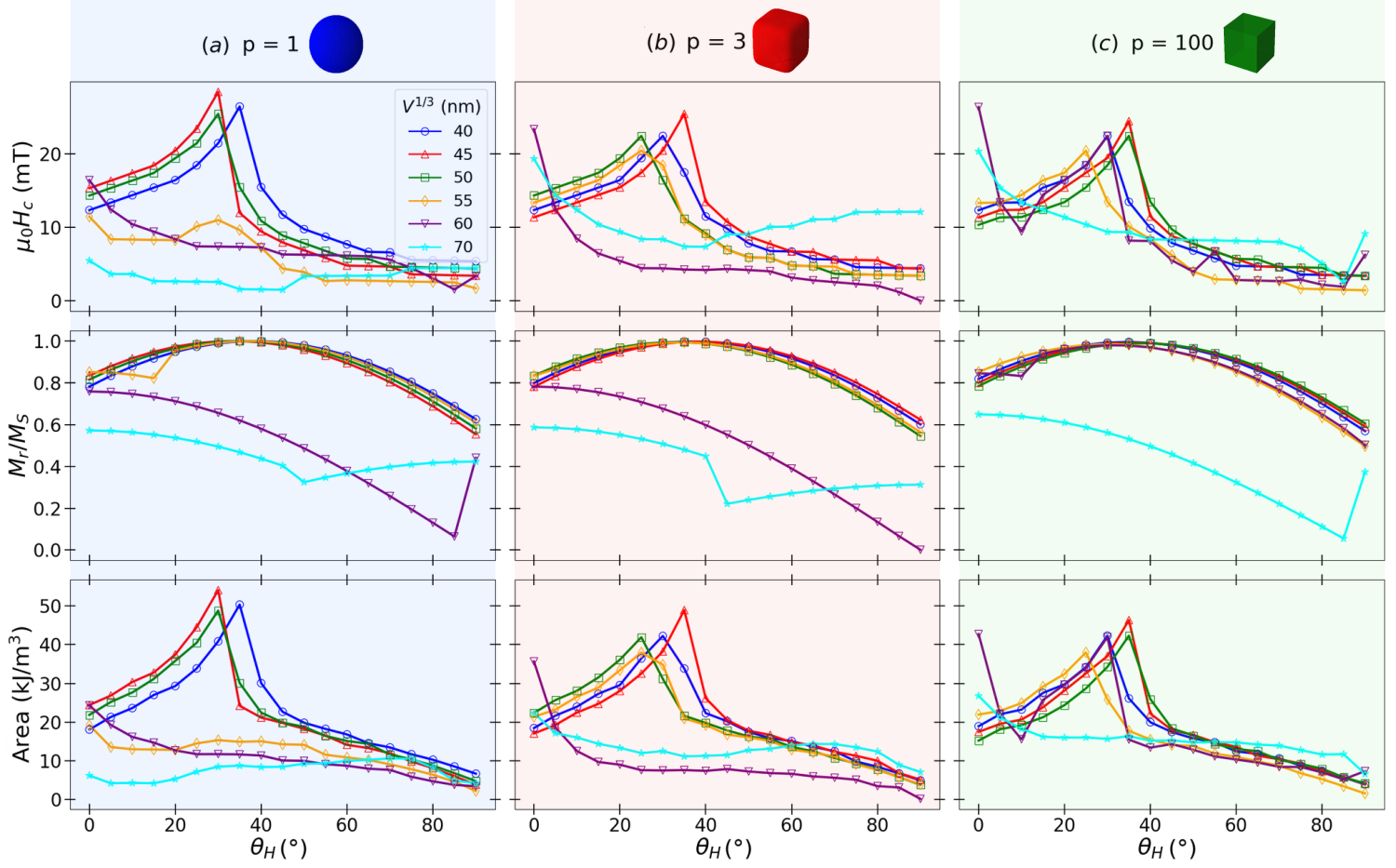}
\caption{Characteristic parameters, $H_c$, $M_r$, and loop area as functions of the angle $\theta_H$, for three shapes: $p=1$ (a), $p=3$ (b), and $p=100$ (c), all with an axial ratio $r = 1.1$, and varying the nanoparticle size $V^{1/3}$. For visual clarity, data points are connected by lines, although this interpolation may not represent the exact trend at intermediate angles.}
\label{critical}
\end{figure*}

\begin{figure*}[!ptb]
\centering
\includegraphics[width=1.0\textwidth]{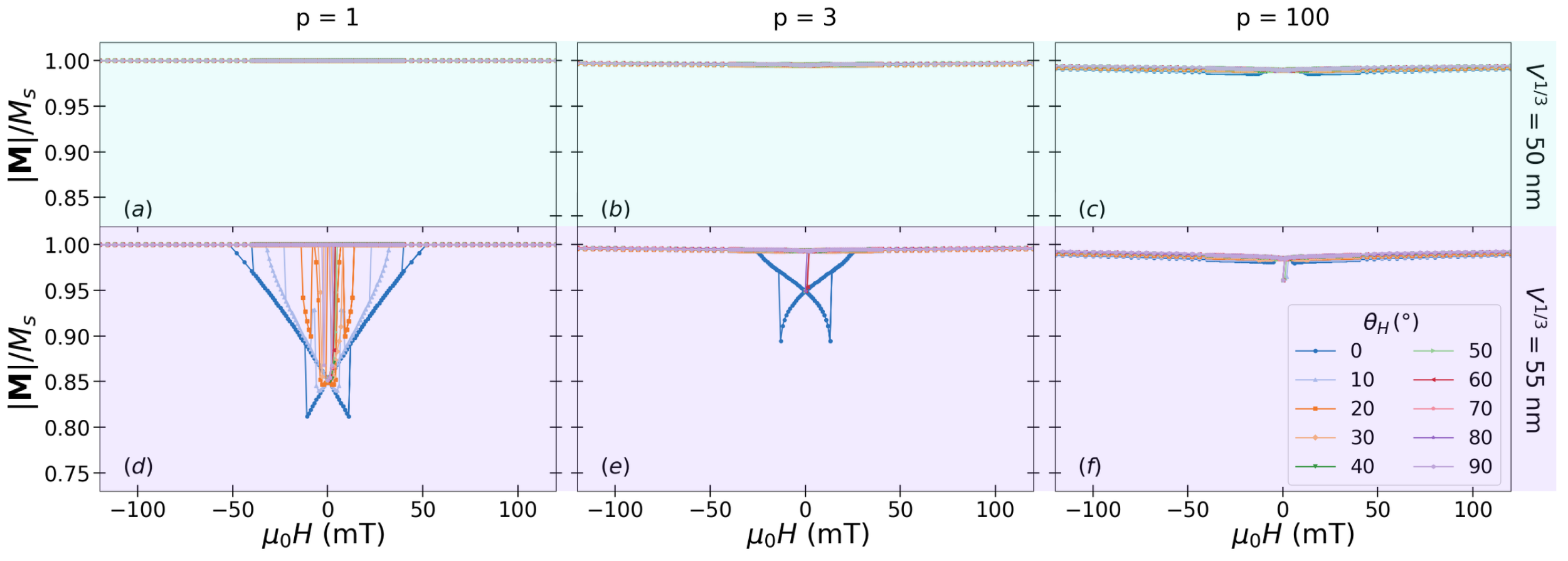}
\caption{Variation of the normalized magnetization modulus $|\textbf{M}|/M_s$ along the hysteresis cycle 
for MNPs with $V^{1/3} = 50$ nm (top row) and $V^{1/3} = 55$ nm (bottom row). 
Each column corresponds to a different shape: (a,d) $p=1$ (sphere), (b,e) $p=3$, and (c,f) $p=100$ (cube).}
\label{modulus}
\end{figure*}

Fig.~\ref{critical} shows clear changes in the trends, most notably the progressive disappearance of the $H_c$ peak around $30^\circ$.
These variations are strongly influenced by the effective particle size. For spherical particles ($p=1$), this inflection becomes evident near $V^{1/3}=55$ nm. 
For $p=3$, deviations appear already at 55 nm and become more pronounced at 60 nm, whereas for $p=100$ the transition occurs around $V^{1/3}=70$ nm. 
Although these changes do not by themselves demonstrate a breakdown of coherent rotation, they do indicate the onset of a size-driven transition between distinct magnetic regimes.

To assess whether the observed trends in $H_c$ are associated with the onset of nonuniform magnetization, we analyze the evolution of the total magnetization modulus $|\textbf{M}|$ throughout the hysteresis cycle. In the coherent (uniform) regime, $|\textbf{M}|/M_s$ remains close to unity at all fields. A systematic reduction of this quantity below unity signals a loss of uniform alignment of the local magnetic moments and the emergence of nonuniform magnetization states.
Figure~\ref{modulus} shows the magnetization modulus for MNPs with sizes $V^{1/3} = 50$ nm (top row) and $V^{1/3} = 55$ nm (bottom row). 

Overall, the results indicate that $V^{1/3} = 55$ nm defines a threshold above which deviations from uniform magnetization become evident. This transition is more pronounced for more rounded particle shapes. For $V^{1/3} = 50$ nm (top row of Fig.~\ref{modulus}), the system still exhibits a quasi-uniform magnetization, with $|\textbf{M}|/M_s \approx 1$, justifying the restriction of the detailed analysis in the main text to this size range, where the macrospin approximation is expected to remain valid.

In our previous work performed in the absence of an external field \cite{Lopez-Vazquez_JMMM_2026}, the critical volume $V_c$ was found to be approximately $49$ nm for spherical MNPs ($p=1$) and $53$ nm for cubic ones ($p=100$), for an axial ratio $r=1.1$. The present results, obtained under an applied field, are very similar, indicating that the field does not significantly modify the critical size $V_c$.

\section{Angular dependence of the hysteresis loops:additional results}\label{ap:4}

\begin{figure*}[!ptb]
\centering
\includegraphics[width=0.9\textwidth]{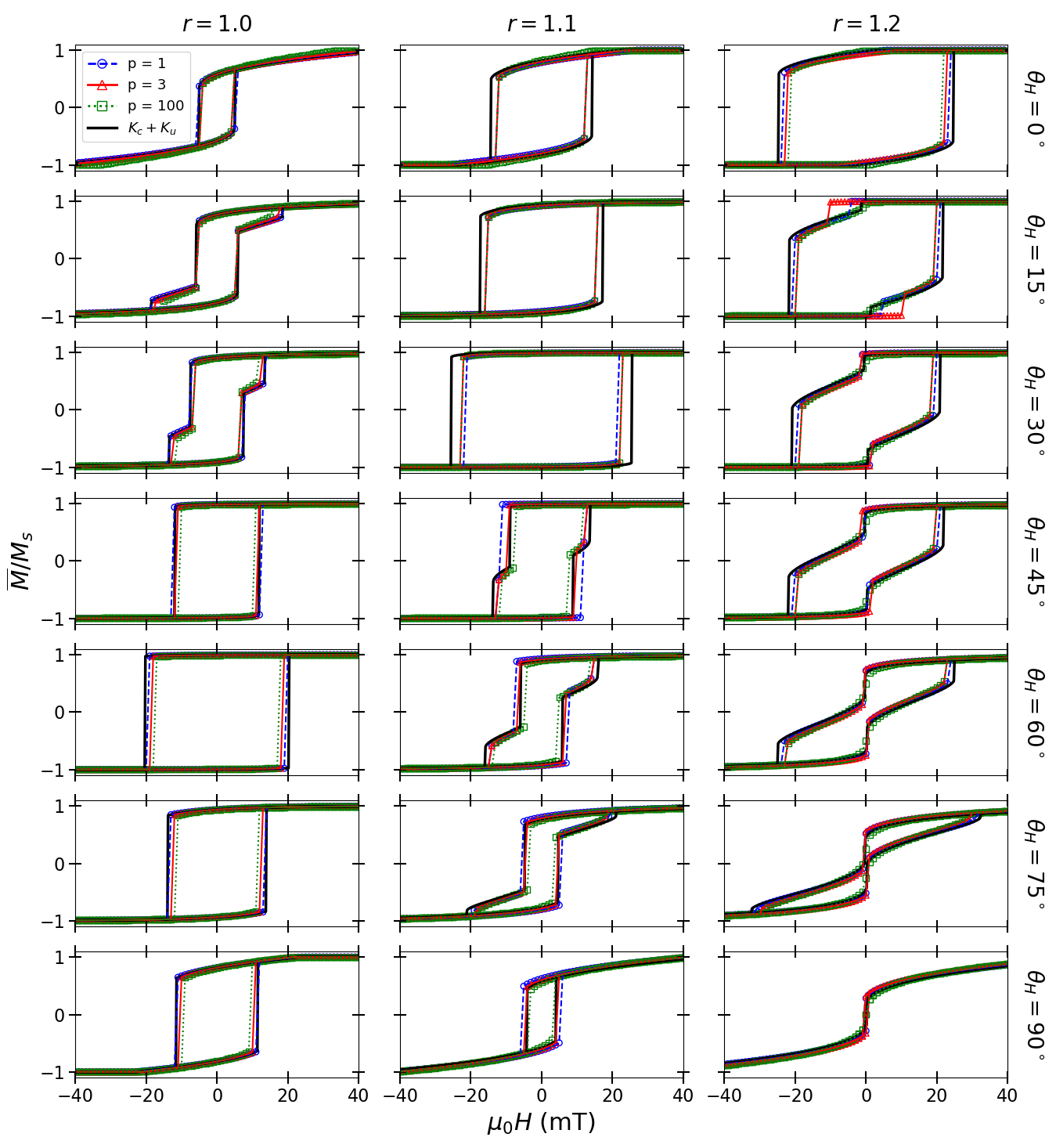}
\caption{Angular dependence of the hysteresis loops obtained from micromagnetic simulations for superellipsoidal nanoparticles with different shape parameters ($p = 1, 3, 100$) and axial ratios ($r = 1.0$, $1.1$, and $1.2$), all with $V^{1/3}=40$ nm. Each row displays the hysteresis loops for different field directions $\theta_H$.}
\label{fig:appendix_angular}
\end{figure*}

% Figure~\ref{fig:appendix_angular} provides the angular dependence of the hysteresis loops for different particle shapes and axial ratios considered in this work. The main conclusion is that the changes in loop morphology are governed primarily by the axial ratio $r$, whereas the dependence on the shape parameter $p$ is comparatively weak. In practice, for a fixed value of $r$, the loops obtained for $p=1$, $3$, and $100$ remain very similar, reinforcing the idea that particle shape plays only a secondary role in the quasi-uniform regime. A more detailed inspection of the angular evolution shows that the hysteresis loop shape changes systematically with $\theta_H$ and depends strongly on elongation. 

Figure~\ref{fig:appendix_angular} shows the angular dependence of the hysteresis loops for the different particle shapes and axial ratios considered in this work. The main conclusion is that the loop morphology is governed primarily by the axial ratio \(r\), whereas the dependence on the shape parameter \(p\) is comparatively weak. In practice, for a fixed value of \(r\), the loops obtained for \(p=1\), \(3\), and \(100\) remain very similar, confirming that particle shape plays only a secondary role within the quasi-uniform regime. By contrast, the loop shape evolves systematically with the field angle \(\theta_H\) and depends strongly on particle elongation.

For $r = 1.0$, where only cubic magnetocrystalline anisotropy is present, the angular dependence departs significantly from the uniaxial SW picture. At small $\theta_H$, the loops present intermediate steps which progressively disappear until $\theta_H\sim 40^{\circ}$. For larger angles, the loops become more square, with an increase in $H_c$ up to $\theta_{min}$, beyond which $H_c$ decreases, as shown in Fig.~\ref{fig:appendix_angular}.
This behavior reflects the presence of multiple equivalent easy axes and competing local minima in the anisotropy energy landscape; the switching fields result from a sequence of instabilities associated with different crystallographic directions.

For finite elongations ($r=1.1, 1.2$), the addition of a shape-induced uniaxial anisotropy term $K_u$ progressively lifts the degeneracy of the cubic energy landscape and introduces a dominant easy axis. 
For $r= 1.1$, the loops remain nearly square for $\theta_H<\theta_c\sim 33^{\circ}$, and then develop intermediate steps whose amplitude increases as $H_c$ decreases. These features indicate that the cubic contribution still perturbs the effective energy landscape.

For larger elongation ($r=1.2$), the uniaxial shape anisotropy dominates over the cubic term. However, the maximum $H_c$ still occurs at $H_c\sim 10^{\circ}$ rather than at $0^{\circ}$, as expected for the $K_u$-only case. Moreover, the steps that appear beyond this maximum occur near $H=0$ and increase in magnitude with $\theta_H$, eventually leading to butterfly-like loops for $\theta_H\gtrsim 75^{\circ}$, where $H_c$ vanishes.
These features are fully consistent with the angular dependence of $H_c$ discussed in Fig.~\ref{critical}, where the persistence of a shifted maximum and the suppression of non-monotonic behavior with increasing $r$ reflect the progressive crossover from a regime dominated by cubic anisotropy to one that approaches an effective uniaxial response.

\section{Comparison of $M_r$ values}
\label{Mr}

To complement the analysis presented in the main text, we evaluated the deviation between the remanence values obtained from micromagnetic simulations and those predicted by the $K_c + K_u$ model. The same MAPE definition introduced in Eq.~\ref{eq:mape} was used, but now applied to the remanence $M_r$:
\begin{equation}
\Delta M_r = 100 \times \frac{1}{n} \sum_{i=1}^{n} 
\left| \frac{M_{r,i}^{\text{micro}} - M_{r,i}^{\text{macro}}}{M_{r,i}^{\text{macro}}} \right|.
\label{eq:mape_Mr}
\end{equation}

\begin{figure}[!thp]
    \centering
    \includegraphics[width=\columnwidth]{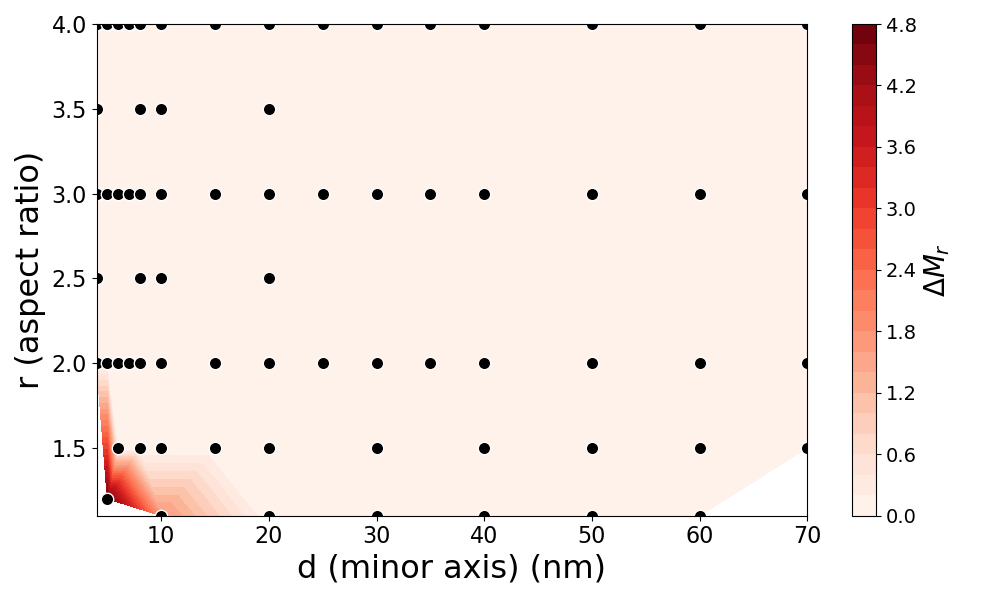}
    \caption{Mean deviation \(\Delta M_r\) between the micromagnetic results and the $K_c + K_u$ model as a function of the minor axis size \(d\) and the axial ratio \(r\).}
    \label{fig:diagrama_Mr}
\end{figure}
As shown in Fig.~\ref{fig:diagrama_Mr}, the deviation in remanence, $\Delta M_r$, remains below 2\% across nearly the entire range of particle sizes and elongations analyzed. This demonstrates that the remanent magnetization is essentially preserved even when the magnetization begins to lose coherence. In contrast, the coercive field exhibits a much stronger dependence on the onset of non-uniform magnetization states.

Therefore, the coercive field $H_c$ serves as a more sensitive indicator of the emergence of non-uniform reversal processes, which supports its use as the primary parameter for constructing the phase diagram presented in the main text.

% \nocite{*}

\bibliographystyle{apsrev4-2}
\bibliography{New}% Produces the bibliography via BibTeX.

\end{document}